\documentclass[pre,aps,floats,floatfix,superscriptaddress,nofootinbib,citeautoscript,twocolumn,showpacs]{revtex4}
\usepackage{epsfig}
\usepackage{graphicx}
\usepackage{amssymb,amsmath}
\usepackage{float}
\usepackage{version}
\usepackage{hyperref}
\usepackage{times}
\bibpunct{[}{]}{,}{n}{}{,}

\newcommand{\dd}{\textrm{d}}

\newcommand{\ee}{\text{e}}
\newcommand{\bv}{{\bf v}}

\newcommand{\br}{{\bf r}}

\newcommand{\bs}{\hat{\boldsymbol \sigma}}
\newcommand{\bsig}{\hat{\boldsymbol \sigma}}

\newcommand{\bG}{{\boldsymbol \Gamma}}
\newcommand{\bg}{{\bf g}}

\newcommand{\vt}{{\tilde v}}
\newcommand{\tv}{{\tilde v}}

\newcommand{\tf}{{\tilde f}}

\newcommand{\tP}{{\widetilde P}}

\newcommand{\vds}{{v_{2 \sigma}}}
\newcommand{\vdst}{{\tilde{v}_{2 \sigma}}}

\newcommand{\cN}{{\cal N}}

\newcommand{\cO}{{\cal O}}

\newcommand{\hP}{{\hat P}}

\newcommand{\hf}{{\hat f}}
\newcommand{\Mvariable}{{}}
\newcommand{\Mfunction}{{}}
\newcommand{\p}{\partial}
\graphicspath{{figures/eps/}}

\begin{document}
\title{Collisional statistics of the hard-sphere gas} 

\date{\today} 

\author{Paolo Visco} 

\affiliation{Universit\'e Paris-Sud, LPTMS, UMR 8626, Orsay Cedex,
F-91405 and CNRS, Orsay, F-91405}

\affiliation{SUPA, School of Physics, University of Edinburgh,
Mayfield Road, Edinburgh, EH9 3JZ, UK}

\author{Fr\'ed\'eric van Wijland}

\affiliation{ Laboratoire Mati\`ere et Syst\`emes Complexes (CNRS UMR
  7057), Universit\'e Denis Diderot (Paris VII), 10 rue Alice Domon et
  L\'eonie Duquet, 75205 Paris cedex 13, France}

\author{Emmanuel Trizac} 

\affiliation{Universit\'e Paris-Sud, LPTMS, UMR 8626, Orsay Cedex,
   F-91405 and CNRS, Orsay, F-91405}

\begin{abstract}
  We investigate the probability distribution function of the free
  flight time and of the number of collisions in a hard sphere gas at
  equilibrium. At variance with naive expectation, the latter quantity
  does not follow Poissonian statistics, even in the dilute limit
  which is the focus of the present analysis. The corresponding
  deviations are addressed both numerically and analytically.  In
  writing an equation for the generating function of the cumulants of
  the number of collisions, we came across a perfect mapping between
  our problem and a previously introduced model: the probabilistic
  ballistic annihilation process [Coppex {\it et al.}, Phys. Rev. E
  {\bf 69} 11303 (2004)]. We exploit this analogy to construct a
  Monte-Carlo like algorithm able to investigate the asymptotically
  large time behavior of the collisional statistics within a
  reasonable computational time. In addition, our predictions are
  confronted against the results of Molecular Dynamics simulations and
  Direct Simulation Monte Carlo technique. An excellent agreement is
  reported.
\end{abstract}

\pacs{05.40.-a, 51.10.+y, 02.50.-r}
\maketitle

\section{Introduction}
Despite the great interest that the hard sphere gas has triggered
since the early days of statistical physics, there are still, to this
day, simple questions which have not been deeply investigated yet. In
this paper we address one such question: what is the probability that
a tagged particle suffers a given number of collisions in a time $t$?
Apart from the simplicity of the question, relevant in its own right,
the collisional statistics of the hard sphere gas turns out to be
useful in several purposes, such as the estimation of transport
coefficients or the characterization of the dissipation in the closely
related granular gases.

In a low density hard sphere gas at equilibrium, the velocities of two
colliding particles are uncorrelated just before the collision: this
is the molecular chaos (strosszahlansatz) statement.  This remark
could naively lead to the conclusion that collisions are uncorrelated
random events, implying therefore that the number of collisions is
simply a Poisson random variable. However, even if molecular chaos is
exactly verified, the collision number is {\it not a Poisson random
variable}.  The non-Poissonian nature of collisional statistics has
already been noticed in the literature (see e.g.  \cite{lue}) but to
our knowledge, has resisted analytical investigations.  It is our
purpose here to fill this gap.

The reason for the non-Poissonian behavior alluded to above is that
the probability for a collision to take place depends on the
scattering cross section of the colliding pair (which depends on the
relative velocity $\bg$ of the pair).  Namely, for the hard sphere
gas, the probability of a collision behaves roughly as $\bg \cdot
\bsig \Theta(\bg \cdot \bsig)$, where $\bsig$ is a unitary vector
joining the centers of mass of the two particles at contact (directed
along the apse line), and $\Theta$ is the Heaviside step function.
For a gas made up of particles interacting through a potential other
than the hard-sphere potential, the probability of having a
``collision'' is of course different, which affects the distribution
of the number of collisions.  Of particular interest is the gas of
particle interacting through a pair potential $V(\br) \sim 1/r^{2
(d-1)}$, where $d$ is the space dimension. Such particles are known as
Maxwell molecules \cite{Ernst} and lead to a velocity independent
probability of having a collision. Within this model, it then appears
that the collisions are truly uncorrelated random events, and hence
that the number of collision is a Poisson random variable.

In the next section, the free flight time distribution of a hard
sphere gas is addressed, together with the distribution of free path
lengths. These results pertain to $\cN=0$ collision properties.  In
section \ref{sec:nbcoll}, the analysis is extended to consider the
full probability $P(\cN,t)$ encoding the number of collisions
statistics. The large time behavior and a complementary perturbative
treatment are worked out analytically.  Explicit expressions are
obtained for the cumulants of the number of collisions $\cN$.  In
order to put the theoretical predictions to the test, three types of
numerical simulations are performed.  The first two, Molecular
Dynamics and Direct Simulation Monte Carlo (DSMC) are routinely
employed in the field and beyond. The third type, of the Monte Carlo
family, is discussed in \ref{sec:simulation}, and specifically devised
to solve the particular problem at hand. It relies on a
reinterpretation of the eigenvalue equation for the generating
function for the cumulants of $\cN$, in terms of a population dynamics
with cloning and annihilation.  A Markov chain with the desired
properties in then constructed, which allows for a direct measure of
several key quantities involved in the analytical treatment.  In doing
so, we uncover a fruitful mapping with the probabilistic ballistic
annihilation model proposed in Ref \cite{coppex04}.  The three
numerical methods provide results that are in excellent agreement with
the analytical predictions.  Preliminary accounts of part of this work
has appeared in \cite{VvWT}, where the numerical aspect of the work
was limited to its Molecular Dynamics content, and where the theory is
restricted to the zeroth order of the treatment put forward here.

\section{Free flights time distribution}
\subsection{General results}
\begin{figure}
  \includegraphics[width=0.46\textwidth,clip=true]{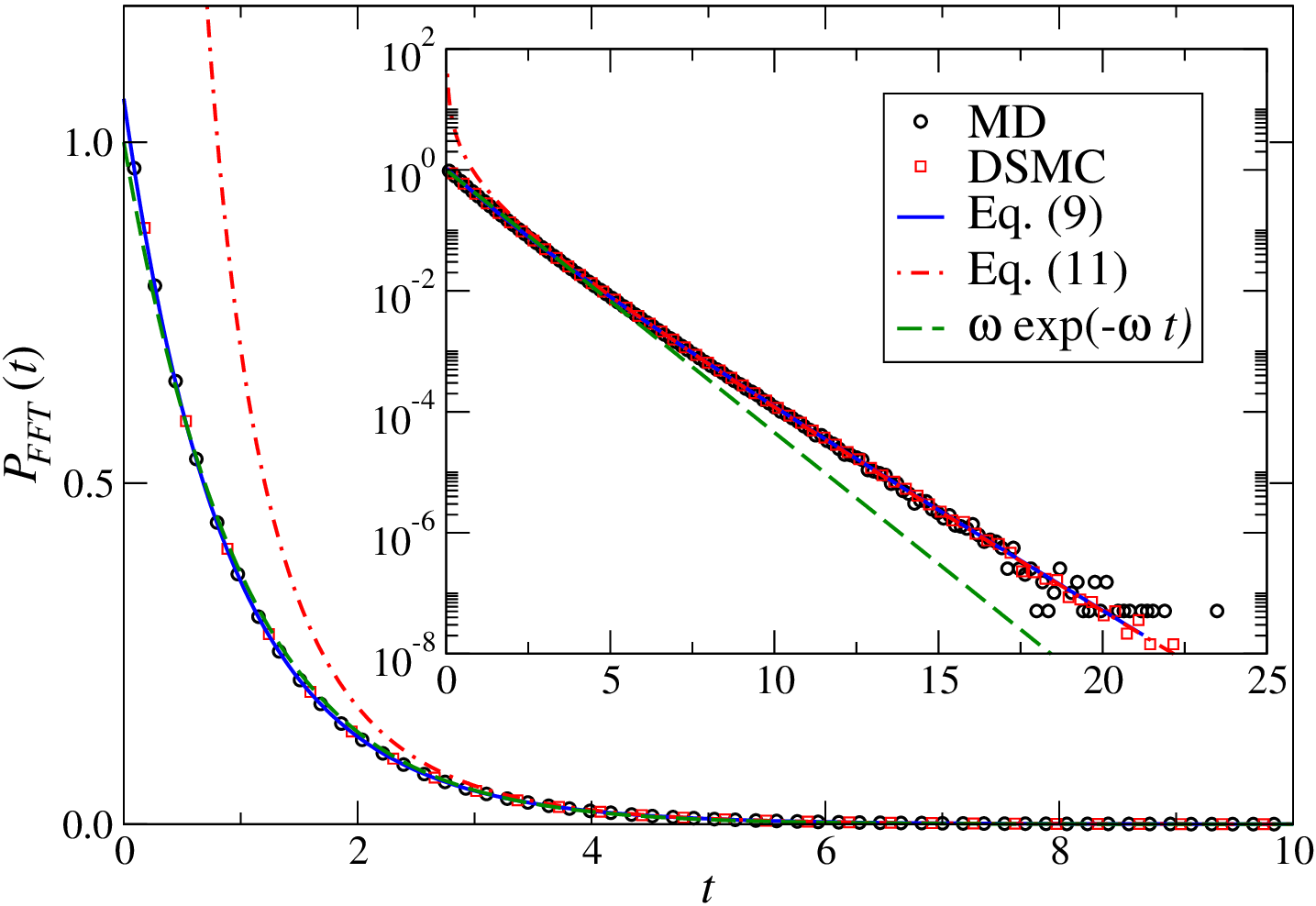}
  \caption{\label{fig:fft}(Color online) Free flights time
    distribution of a hard disc gas. The circles and squares
    correspond to the results of Molecular Dynamics (MD) and Direct
    Simulation Monte-Carlo (DSMC) methods. MD simulations in all this
    work are shown at density $\rho=0.01 \sigma^{-2}$.  The full line
    is the numerical integration of Eq. (\ref{eq:fft}). The dot-dashed
    line shows the asymptotic large time behavior of
    Eq. (\ref{eq:fftlt}), and the dashed line is the exponential
    distribution, associated to the Poisson distribution. Time is
    measured in units of the mean free time ($\omega=1$).}
\end{figure}

We consider a hard sphere gas in $d$ dimensions composed of $N$
particles of equal mass $m$ and equal diameter $\sigma$. The gas is
thermalized at some temperature $T_0$ in a homogeneous state of
(constant) density $\rho$, so that the one point distribution function
of the system is a Gaussian:
\begin{equation}
  \phi(\br,\bv)\equiv \phi(\bv)= \rho (2 \pi T_0)^{-d/2} 
  \exp \left(- \frac{v^2}{2 T_0} \right)
  \,\,.
\end{equation}
If one decides to follow the evolution of one particle among the $N$,
then the velocity probability distribution function (pdf) of that
particle will evolve according to the homogeneous linear Boltzmann
equation:
\begin{equation}
\label{linboltz}
\frac{\p}{\p t}  f(\bv_1,t)=
\frac{1}{\ell} \int \!\!
\dd \bv_2 \int' \!\!\!\! \dd \bs \  
\bv_{12} \cdot
\bs  \left[f_1^{**} \phi_2^{**} - 
 f_1  \phi_2 \right]\,\,,
\end{equation}
where $\ell=(\sigma^{d-1} \rho \chi)^{-1}$ is a length proportional to
the mean free path denoted $l$ below, and $\bv_{12}=\bv_1-\bv_2$ .
The factor $\chi$ is the pair correlation function at contact and is
the so called Enskog correction factor \cite{enskog}. In the dilute
limit where the molecular chaos assumption is justified, it tends to
unity.  Moreover, the primed integral means that the integration has
to be performed in the domain for which $\bv_{12} \cdot \bsig
>0$. Finally, we are employing the short-hand notation $f^{**}_i
\equiv f(\bv_i^{**},t)$, where the two star superscript refers to the
precollisional velocity:
\begin{equation}
  \bv_1^{**}=\bv_1 - (\bv_{12} \cdot \bsig) \bsig\,\,,\qquad 
  \bv_2^{**}=\bv_2 + (\bv_{12} \cdot \bsig) \bsig\,\,.
\end{equation}
In this framework the evolution of our tagged particle is exactly a
Markov process. The probability of hopping from a state with velocity
$\bv$ to another state of velocity in a narrow interval $\dd \bv'$
around $\bv'$ and in a time interval $\dd t$ is given by $W(\bv'|\bv)
\dd \bv' \dd t$, where
\begin{multline}
W(\bv' | \bv)=\frac{1}{\ell} |\bv'-\bv|^{2-d} (2 \pi T_0)^{-1/2} \exp
\Bigg\{- \frac{1}{2 T_0} \times \\ \left[v'^2 - v^2 + \bv \cdot \left(
\frac{\bv'-\bv}{|\bv'-\bv|} \right)\right]\Bigg\}\,\,
\end{multline}
is the transition rate density per unit time (see {\it e.g.}
\cite{puglisi05}). For any general Markov process, the probability of
leaving a given configuration $\bG$ in a time interval $\dd t$ is
$r(\bG) \dd t$, where
\begin{equation}
r(\bG) = \int \dd \bG' W(\bG'|\bG)\,\,.
\end{equation}
This is simply the loss term of the Linear Boltzmann equation
(\ref{linboltz}), and it reads for the hard sphere gas:
\begin{multline}
\label{eq:r}
  r(v)= \frac{\omega}{\sqrt{2}}\,
  \left( \frac{{\Mvariable{v}}^2}{d T_0}\,
    \Mfunction{_1F_1}\left(
      \frac{1}{2},1 + \frac{d}{2},
      -\frac{{\Mvariable{v}}^2}{2\,T_0}\right) \right. + \\ \left. 
     \ee^{-\frac{{\Mvariable{v}}^2}{2\,T_0}} 
    \,\Mfunction{_1 F_1} \left(\frac{d-1}{2},\frac{d}{2},
      \frac{{\Mvariable{v}}^2}{2\,T_0}\right) 
  \right)\,\,, 
\end{multline}
where $_1 F_1$ is the confluent hyper-geometric function of the first
kind \cite{abramowitz}, and $\omega$ is the collision frequency of the
gas:
\begin{equation}
  \omega=\int \dd \bv \phi(\bv) r(v) = \frac{2 \pi^{\frac{d-1}{2}}
  \sqrt{T_0}}{\Gamma(d/2)} \,\,.
\label{eq:omegaa}
\end{equation}
From the Markovian property, it follows that the probability of having
a given velocity $\bv$ for a given time $t$ and making a collision
exactly at time $t$ is exponential:
\begin{equation}
P_{FFT}(t|\bv)=r(v) \exp \left(-r(v) t\right)\,\,.
\end{equation}
The probability density of the free flight time follows:
\begin{equation}
\label{eq:fft}
P_{FFT}(t)=\int \dd \bv \phi_{coll}(\bv) P_{FFT}(t|\bv)\,\,.
\end{equation}
Here $\phi_{coll}(\bv)$ is the velocity pdf of the colliding particles
(i.e. obtained sampling the velocity of the particle only on
collision):
\begin{equation}
\label{eq:phicoll}
\phi_{coll} (\bv)= \frac{r(v)}{\omega} \phi(\bv)\,\,,
\end{equation}
where the normalization factor $\omega$ is exactly the collision
frequency of the gas. The free flight time distribution for hard
spheres was already investigated in \cite{wiegel1}, but the authors
used the equilibrium velocity distribution $\phi$ instead of the on
collision velocity distribution $\phi_{coll}$ as a weight in
(\ref{eq:fft}), and hence obtained an incorrect result (this issue is
further commented in Appendix \ref{app:fftwrong}).  Two measurements
of the free flight time distribution in an event driven molecular
dynamic simulation of hard disks (see section \ref{sec:simulation}
for more details) and in a Direct Simulation Monte Carlo (DSMC)
\cite{bird}, are shown in Fig.~\ref{fig:fft}, together with the
expression of Eq.  (\ref{eq:fft}). The agreement is excellent.  The
molecular dynamics simulations are performed at a low density ($\rho
\sigma^2=10^{-2}$) to ensure that molecular chaos holds, whereas the
DSMC technique relies by construction on the molecular chaos
approximation, and can therefore be considered as providing a $\rho
\to 0$ benchmark.  Figure \ref{fig:fft} also provides a comparison
with the exponential law that would hold for a constant $r(v)$, as is
the case for Maxwell molecules. The latter expectation [$\omega
\exp(-\omega t)$] is seen to hold at early times, but significantly
fails at large $t$, a time regime that we investigate in more details
in the next section.

\subsection{Large time and high dimension analysis}

Apart from simplified models (see Appendix \ref{app:vhp}),
it does not seem possible to obtain a simple exact analytical
expression for the free flight time distribution, mainly because of
the particular behavior of the hyper-geometric function involved in
the expression of the collision rate $r(v)$. In order to get a simpler
approximated result, one has to look at limiting cases.
For large times, the integral involved in Eq. (\ref{eq:fft}) can be
estimated through the saddle-point approximation (a.k.a. method of
steepest descent, see e.g. \cite{morse}), yielding:
\begin{equation}
\label{eq:fftlt}
P_{FFT}(t) \sim \exp\left({- \frac{\omega t}{\sqrt{2}}}\right)
\frac{\omega}{2} \left(1-\frac{2}{d} + \frac{\omega t}{\sqrt{2} d}
\right)^{-d/2}\,\,.
\end{equation}
The above function is plotted in Fig. \ref{fig:fft}, and seen to be in
very good agreement with our numerical data provided $t$ is large
enough. Note that expression (\ref{eq:fftlt}) does not provide a
normalized probability (a feature visible on Fig. \ref{fig:fft}), but
only an asymptotic expansion.

Besides, a way to get an explicit expression for the collision rate
$r(v)$ is to investigate the infinite dimension limit. In this case,
the saddle-point approximation yields:
\begin{equation}
\label{rlarged}
r(v) \simeq \frac{\omega}{\sqrt{2}} \sqrt{1+\frac{v^2}{2 T_0 d}}\,\,.
\end{equation}
A detailed derivation of this result is given in Appendix
\ref{app:saddle}.

\subsection{Free path length distribution}
\begin{figure}
  \includegraphics[width=0.46\textwidth,clip=true]{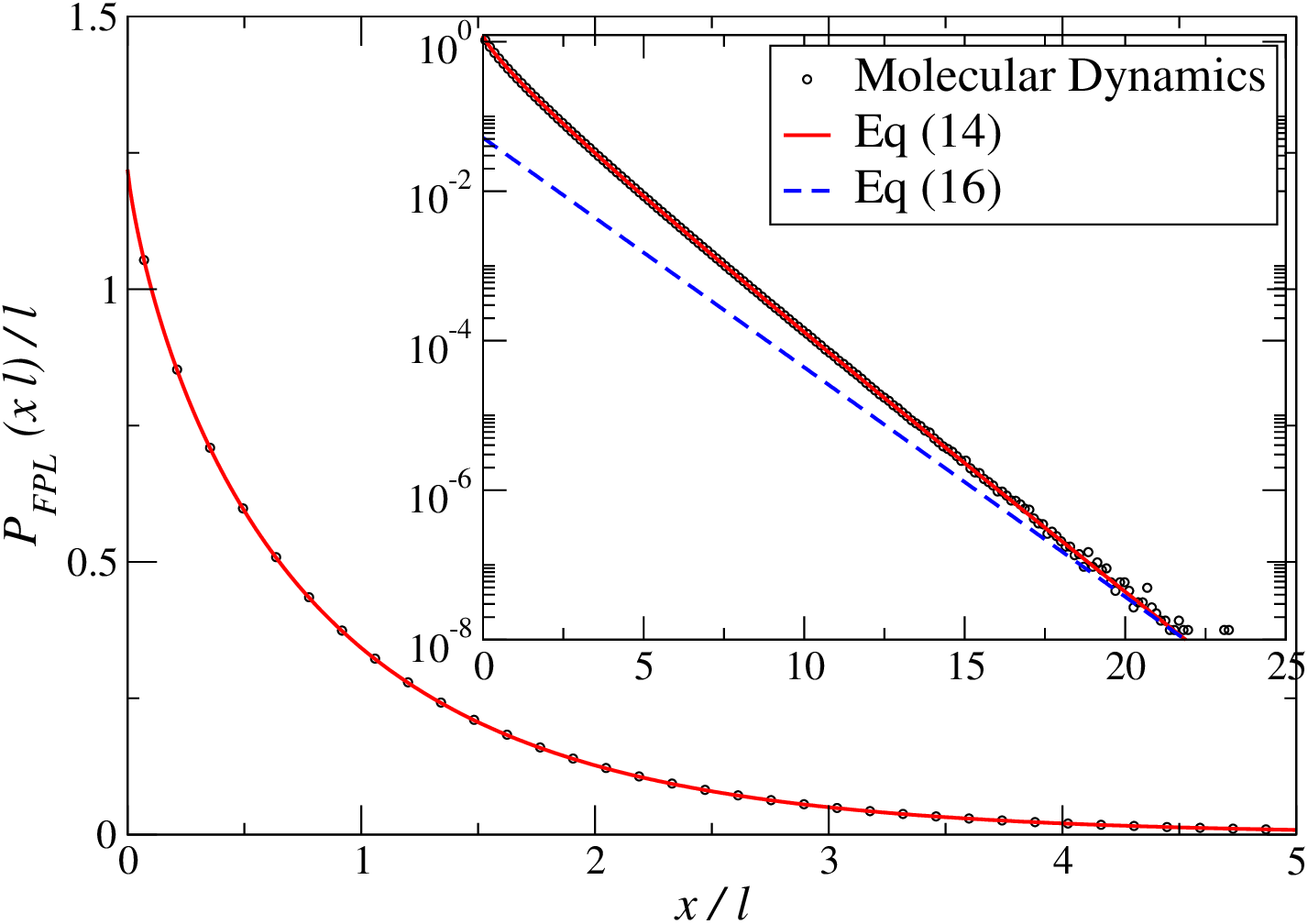}
  \caption{\label{fig:freepath_hs} (Color online) Plot of the free
    path length distribution from MD simulations (at density
    $\rho=0.01 \sigma^{-2}$) and from a numerical integration of
    Eq. (\ref{eq:fpl}). In the inset, the dashed line is the dominant
    exponential term, Eq. (\ref{eq:dom}).%
}
\end{figure}

From the knowledge of the free flight time distribution, it is
straightforward to compute the distribution of the free path
length. In fact, for a particle with a velocity $\bv$, the length
covered in a time $t$ is $x=|\bv| t = v t$. Therefore, the conditional
distribution of free path length is obtained as:
\begin{multline}
  P_{FPL}(x|\bv)= \frac{1}{v} P_{FFT}\left( \left.\frac{x}{v}\right|
    \bv \right) \\ = \frac{r(v)}{v} \exp \left(-\frac{r(v)}{v}
    x\right)\,\,.
\end{multline}
The free path length distribution can be obtained averaging the above
expression with the weight $\phi_{coll}$:
\begin{equation}
\label{eq:fpl}
P_{FPL}(x)=\int \dd \bv \phi_{coll}(\bv) P_{FPL}(x|\bv)\,\,.
\end{equation}
Although the conditional probability of free path length is very
similar to the conditional probability of free flight time, the step
of averaging over the collisional velocity distribution may
drastically change the shape of the distribution. A case worthy of
attention is discussed in Appendix \ref{app:fplmaxwell}, for Maxwell
molecules: the free flight time distribution is a pure exponential,
while the free path length distribution has a stretched exponential
decay. In the case of hard particles, the dominant large path behavior
remains exponential, as for the free flight time. Nevertheless, the
leading exponential prefactor changes. The free flight time
distribution is dominated by $\ee^{-r(0) t}$, where $r(0)$ is the
minimum of the function $r(v)$. In the case of free path length, the
dominant exponential behavior is $\ee^{A x}$, where $A$ is the minimum
of $r(v)/v$, which is obtained in the limit $v \to \infty$:
\begin{equation}
  A=\lim_{v \to \infty} \frac{r(v)}{v}=\frac{1}{l \sqrt{2}}=
    \frac{{\pi }^ { \frac{d-1}{2} }}{\ell \,\Mfunction{\Gamma}(
    \frac{d+1}{2})} \,\,,
\end{equation}
where $l=\omega/\langle v \rangle$ is the mean free path. This yields:
\begin{equation}
P_{FPL}(x) \sim \ee^{\frac{x}{\sqrt{2} l}}\,\,,
\label{eq:dom}
\end{equation}
(here the $\sim$ should be understood as an equivalence between the
logarithms in the large $x$ limit).  Unfortunately, this expression
only gives the behavior of the dominating exponential term, which
becomes visible at a length scales significantly larger than the mean
free path. Hence, if one analyzes the result of MD simulations as in
Fig.  \ref{fig:freepath_hs}, one sees that there are sub-leading
(algebraic) terms in the asymptotic behavior, which still play a role
and that are responsible for the mismatch between Eq. (\ref{eq:dom})
and MD data in Fig. \ref{fig:freepath_hs}. A similar feature holds for
the free flight time distribution: upon neglecting the algebraic
sub-dominant correction in Eq. (\ref{eq:fftlt}), the agreement
displayed at late times in the inset of Fig.  \ref{fig:fft} would be
spoiled.

\section{The number of collisions}
\label{sec:nbcoll}

\subsection{Uncorrelated approach}
After having shown how the probability of the time between two
subsequent collisions behaves, our interest goes to the probability of
the sum of many such times, and in particular, the probability of
having $\cN$ collisions in a time $t$. The time interval $t$ in which
the $\cN$ collisions take place can be decomposed in a sum of $\cN+1$
(correlated) times:
\begin{equation}
t=t_{in} + t_1 + \dots + t_{\cN-1} + t_f\,\,.
\end{equation}
We are considering that a given particle begins a trajectory with a
given initial velocity, and then waits a time $t_{in}$ before it makes
a collision and gets a new velocity. During a time $t_1$, it then
flies freely, before colliding again etc.  The successive behavior of
the particle's trajectory will be a sequence of collisions, spaced by
other free flight times $t_2$, $t_3$, $\dots, t_{\cN-1}$, until the
last $\cN$-th collision, which will be followed by a final time
interval $t_f$ which is not ended with a collision. The free flight
times $t_1$ to $t_{\cN-1}$ are of course distributed following the
probability $P_{FFT}$ given by Eq.  (\ref{eq:fft}). We will denote by
$P_{in}$ and $P_f$ the pdf of $t_{in}$ and $t_f$, although their
precise statistics are irrelevant for our purposes, as becomes clear
below.

If one would assume these $\cN+1$ times to be uncorrelated, then the
probability $P(\cN,t)$ of having $\cN$ collisions in a time $t$ could
be deduced from the knowledge of the free flight time probability as
follows. We first introduce the Laplace transform:
\begin{equation}
\tP(z) = \int \! \dd t \,\, \ee^{- z t } P(t)\,\,,
\end{equation}
where $P$ can be either $P_{FFT}$, $P_{in}$, or $P_f$.
The transform of $P(\cN,t)$ can hence be expressed as:
\begin{equation}
\tP(\cN, z) = \tP_{in}(z)\,\, \tP_{FFT}(z)^{\cN-1} \,\,\tP_f(z)\,\,.
\end{equation}
Of course, since $P_{FFT}$ cannot be computed analytically, a
closed form expression for $\tP(\cN,z)$ is not available. 
One interesting limit to analyze 
is that of large times. 
To get an asymptotic form of the following integral
\begin{equation}
\label{eq:int}
P(\cN,t)= \int \dd z \,\, \ee^{z t} \tP_{in}(z)\,\, 
\tP_{FFT}(z)^{\cN-1} \,\,\tP_f(z)\,\,,
\end{equation} 
we make the remark that the number of collisions increases with time
linearly on average, and we define a time intensive counterpart to
$\cN$, a fluctuating collision rate, which we denote by $n \equiv
\cN/t$. The integral of Eq. (\ref{eq:int}) then reads:
\begin{equation}
  P(\cN,t)= \int \dd z \,\, \exp \left( t (z + n \ln(\tP_{FFT}(z))) 
    +\cO (1) \right)\,\,.
\end{equation} 
Finally, when $t \to \infty$, the above probability behaves as:
\begin{equation}
P(\cN,t) \sim \ee^{\pi(n) t}\,\,,
\end{equation}
where $\pi(n)$ is a large deviation function, and is related to
$\tP_{FFT}$ through the following relation:
\begin{equation}
\label{piuncorr}
\pi(n) = \min_{z} (z + n \ln(\tP_{FFT}(z))) \,\,.
\end{equation}

\subsection{Large time behavior}
The result of the previous subsection leads to an approximate
evaluation of the large deviation function of the number of collisions
that neglects temporal correlations. We will now investigate this
large deviation function keeping into account these correlations. Let
us define the joint probability $f(\bv,\cN,t)$ of having a velocity
$\bv$ and having suffered $\cN$ collisions up until time $t$. In a
homogeneous state, the time evolution of the above defined probability
is governed by a slightly different form of the linear Boltzmann
equation:
\begin{multline}
  \p_t f(\bv_1,\cN,t)=\int \dd \bv_2 \int' \dd \bs (\bv_{12} \cdot
  \bs) \times \\ \left[f(\bv_1^{**},\cN-1,t) \phi(\bv_2^{**}) -
    f(\bv_1,\cN,t) \phi(\bv_2) \right] \,\,.
\end{multline}
The function $\phi(\bv)$ still represents the one point velocity pdf
of the gas, which is in equilibrium. Moreover, since the particle
whose collisions we are counting has the same velocity pdf, we enforce
that:
\begin{equation}
\sum_{\cN=0}^{\infty} f(\bv,\cN) = \phi(\bv)\,\,.
\end{equation}
It is useful to introduce the generating function of $f$ as:
\begin{equation}
\hf(\bv,\lambda,t) = \sum_{\cN = 0}^{\infty} \ee^{-\lambda \cN} f(\bv,
\cN,t) \,.
\end{equation}
It can be seen that $\hf$ evolves according to
\begin{multline}
\p_t \hf(\bv_1, \lambda,t) = \int \dd \bv_2 \int \dd \bs
\Theta(\bv_{12} \cdot \bs) (\bv_{12} \cdot \bs) \times \\ [
\ee^{-\lambda} f(\bv_1^{**},\lambda,t) \phi(\bv_2^{**}) -
f(\bv_1,\lambda,t) \phi(\bv_2)]\,\,.
\end{multline}
The large time behavior of the solution of the above equation is
dominated by the largest eigenvalue $\mu(\lambda)$ of the evolution
operator:
\begin{multline}
\label{eq:eigenvalue}
\mu(\lambda) \tf(\bv_1,\lambda)=L_{\lambda}\tf(\bv_1,\lambda)=\\ 
\int \dd \bv_2 \int \dd \bs \Theta(\bv_{12}
\cdot \bs)  (\bv_{12} \cdot \bs) \times \\
[ \ee^{-\lambda} \tf(\bv_1^{**},\lambda)
\phi(\bv_2^{**}) - \tf(\bv_1,\lambda) \phi(\bv_2)]\,\,,
\end{multline} 
where $\tf$ denotes the eigenfunction of $L_{\lambda}$ associated with
$\mu$:
\begin{equation}
\hf(\bv, \lambda,t) \,\propto\, e^{\mu(\lambda) t} \tf(\bv, \lambda) .
\end{equation}
Moreover, since $\hP (\lambda) \sim \ee^{\mu(\lambda) t}$, one sees
that $\mu(\lambda)$ is proportional to the cumulant generating
function:
\begin{equation}
\langle \cN^p \rangle_c= t (-1)^p \left. \frac{\p^p \mu}{\p \lambda^p}
\right|_{\lambda=0}\,\,.
\end{equation}
Furthermore, $P(\cN) \sim \int \dd \lambda \ee^{\lambda \cN}
\hP(\lambda)$, and hence, applying the saddle point method for $t \to
\infty$, one has that
\begin{equation}
P(\cN,t) \sim \exp(t \pi(n))\,\,, 
\end{equation}
where $n=\cN/t$ and the large deviation function $\pi$ is related to
$\mu$ through a Legendre transform:
\begin{equation}
\pi(n)=\min_{\lambda} (\mu(\lambda) + \lambda n)\,\,.
\end{equation}

\subsubsection{Large $\lambda$ behavior}
Before trying to solve Eq. (\ref{eq:eigenvalue}) extending the methods
of kinetic theory, we shall first provide exact results, which can be
extracted from the analysis of asymptotically large values of
$\lambda$. The evolution operator appearing in
Eq. (\ref{eq:eigenvalue}) can be cast in the form:
\begin{equation}
L_{\lambda}=L_0 + \ee^{-\lambda} L_1\,\,,
\end{equation}
where
\begin{subequations}
\begin{multline}
L_0 \tf(\bv_1,\lambda) =- \int \dd \bv_2 \int' \dd \bs (\bv_{12} \cdot
\bs) \tf(\bv_1,\lambda) \phi(\bv_2) = \\-r(v_1)
\tf(\bv_1,\lambda)\,\,,
\end{multline}
\begin{equation}
  L_1 \tf(\bv_1,\lambda) =\int \dd \bv_2 \int' \dd \bs (\bv_{12} \cdot
  \bs) \tf(\bv_1^{**},\lambda) \phi(\bv_2^{**}) \,\,.
\end{equation}
\end{subequations}
For large values of $\lambda$, the coefficient $\ee^{-\lambda}$ plays
the role of a small parameter, and therefore the eigenvalue equation
(\ref{eq:eigenvalue}) can be solved in perturbation theory. We will
therefore try to get an expression of $\mu(\lambda)$ for large
$\lambda$ of the form:
\begin{equation}
\mu(\lambda) = \mu^{(0)} + \ee^{-\lambda} \mu^{(1)} 
+ {\cal O} (\ee^{-2 \lambda})\,\,.
\end{equation}
To zero-th order, the largest eigenvalue of $L_{\lambda}$ is given
by the maximum of the function $-r(v)$, whose expression is written in
Eq. (\ref{eq:r}).  The maximum of this function occurs at $v=0$, and
\begin{equation}
\label{r0}
\mu^{(0)}=-r(0)=-\frac{\omega}{\sqrt{2}}\,\,.
\end{equation}
The eigenfunction associated with this eigenvalue is indeed a delta
function, centered in $\bv= {\bf 0}$. In order to get the first order
correction to the eigenvalue (\ref{r0}) one has to project the zero-th
order eigenfunction on the operator proportional to the small
parameter:
\begin{equation}
\mu^{(1)}=\int \dd \bv_1 \, L_1 \delta(\bv_1) =
\frac{\omega}{\sqrt{2}}\,\,.
\end{equation} 
Finally one finds that for large $\lambda$, the largest eigenvalue
$\mu$ behaves like:
\begin{equation}
\label{mu_largelambda}
\mu(\lambda) = \frac{\omega}{\sqrt{2}} (\ee^{-\lambda} -1) + {\cal
O} (\ee^{-2 \lambda})\,\,.
\end{equation}
Hence, the probability of having $\cN$ collisions behaves, for values
of $\cN$ small with respect to its average $\langle\cN\rangle=\omega
t$, as a Poisson distribution with a frequency equal to
$\omega/\sqrt{2}$:
\begin{equation}
P(\cN) \sim \frac{\ee^{-\frac{\omega t}{\sqrt{2}}}}{\cN !}
\left(\frac{\omega t}{\sqrt{2}}\right)^{\cN}\,\,,\qquad \textrm{for
$\cN \ll \omega t$}\,\,.
\label{eq:blabla}
\end{equation}
For such a distribution, the large deviation function $\pi$ easily
follows:
\begin{equation}
\pi(n) =  n - n \log (n \sqrt 2/\omega) -\omega/\sqrt{2}
\label{eq:pinpoisson}
\end{equation} 
The result obtained in Eq. (\ref{eq:blabla}) is compatible with the
large time behavior embodied in Eq. (\ref{eq:fftlt}): the time
derivative of the probability of having $\cN=0$ collisions is of
course (minus) the free flight time distribution function.  We
emphasize that both results (\ref{eq:blabla}) and (\ref{eq:fftlt})
hold for large times.  In addition, we note that the only space
dimension dependence involved is through the collision frequency
$\omega$ [see Eq. (\ref{eq:omegaa})].

\subsubsection{An approximate solution}
In order to get an approximate expression for the largest eigenvalue
$\mu(\lambda)$, we will suppose the associated eigenvector $\tf(\bv,
\lambda)$ to be a Gaussian with a given temperature $T(\lambda)$. We
expect that this approximation will provide accurate results for small
values of $\lambda$, given that for $\lambda=0$, $\tf(\bv,0)$ is
exactly a Gaussian with a temperature $T_0$, but it is not {\it a
priori} a systematic approximation for larger values of
$\lambda$. Projecting the Boltzmann-like equation
(\ref{eq:eigenvalue}) onto the first two velocity moments of the
eigenfunction we want to compute, we are left with two closed
equations for $\mu(\lambda)$ and for $T(\lambda)$:
\begin{equation}
\label{eq:appsol}
\mu (\lambda)=\nu_0\,\,,\qquad \qquad  \qquad 
\mu(\lambda) d T(\lambda) = \nu_2\,\,,
\end{equation}
where the $\nu_n$ are the collisional moments (the expression of the
first ones is given in Appendix \ref{app:coll}):
\begin{multline}
\label{collmom}
\nu_n=\int \dd \bv_1 \int \dd \bv_2 \int' \dd \bs v_1^n (\bv_{12}
\cdot \bs) \times \\\left[\tf(\bv_1^{**},\lambda) \phi(\bv_2^{**}) -
\tf(\bv_1,\lambda) \phi(\bv_2) \right]\,\,.
\end{multline}
Solving simultaneously Eqs. (\ref{eq:appsol}), one obtains:
\begin{subequations}
\label{mu_noa2}
\begin{equation}
\mu(\lambda)=- \frac{\omega}{\sqrt{2}} (1-\ee^{-\lambda})
\sqrt{1+\frac{T(\lambda)}{T_0}}\,\,,
\end{equation}
\begin{equation}
T(\lambda)=\frac{\sqrt{2} T_0}{\sqrt{1+\ee^{\lambda}}}\,\,.
\end{equation}
\end{subequations}
One may notice that this result satisfies some of the previous
requirements obtained from the asymptotic large $\lambda$ analysis. In
particular $\mu(\infty)=-\omega/\sqrt{2}$, as in the previous
subsection.  Moreover, the fictitious temperature $T(\lambda)$
vanishes for infinite $\lambda$, meaning that the eigenfunction
associated with $\mu$ does indeed tend towards a delta
function. Nevertheless, while these features concerning the zero-th
order perturbation results of the previous section are fulfilled, the
behavior of Eq. (\ref{mu_noa2}) is different at the next order. In
fact, one can see from Eq. (\ref{mu_noa2}) that, for $\lambda \to
\infty$:
\begin{equation}
\mu(\lambda) \sim -\frac{\omega}{\sqrt{2}} - 
\omega \, \frac{\sqrt{\ee^{- \lambda}}}{2} +
{\cal O} (\ee^{-\lambda})\,\,,
\end{equation} 
to be compared with Eq. (\ref{mu_largelambda}). This is a deficiency
of the Gaussian approximation.

\begin{figure}
  \includegraphics[width=0.46\textwidth,clip=true]{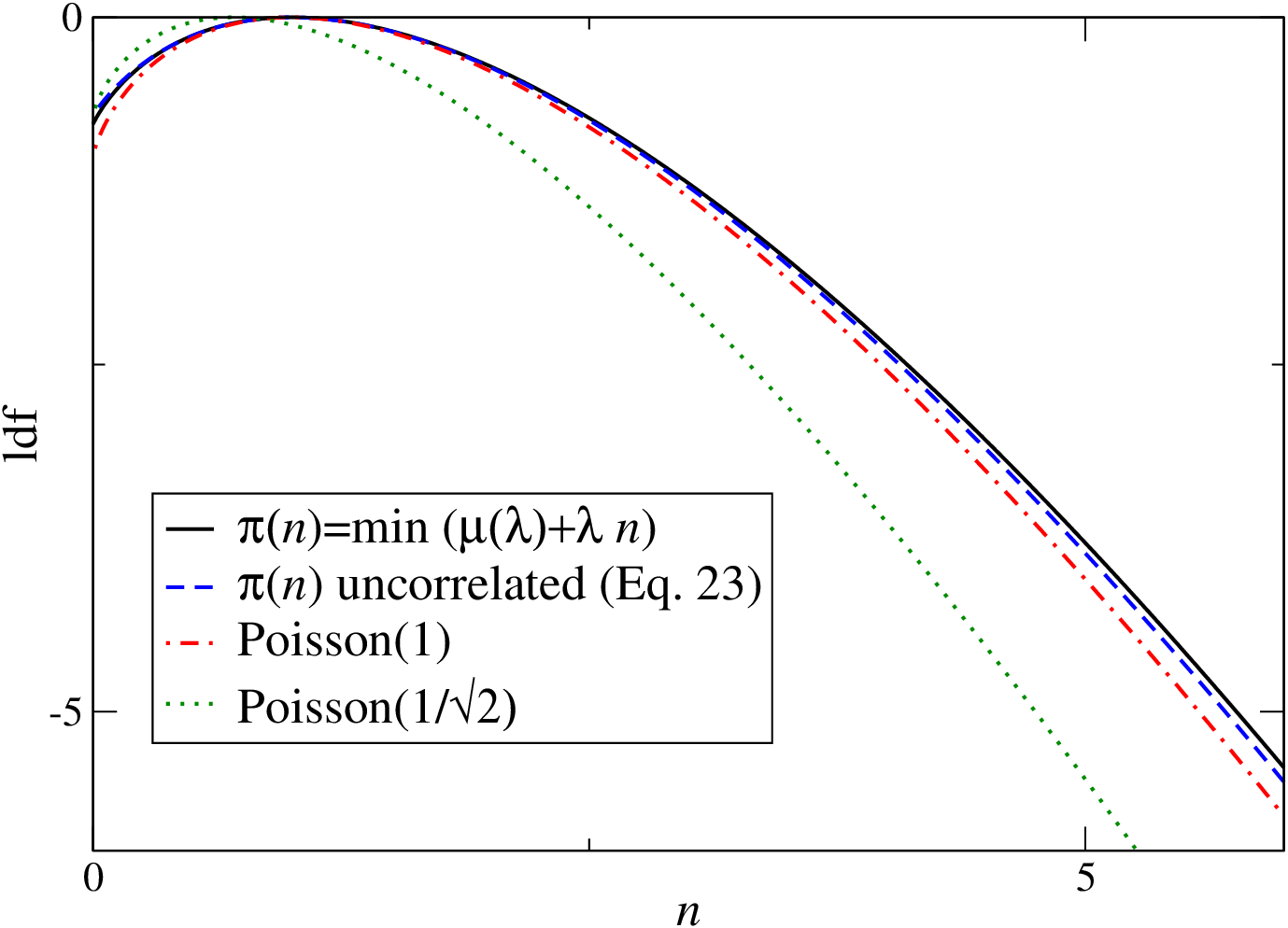}
  \caption{\label{pi_noa2} (Color online) Plot of $\pi(n)$ obtained in
    several fashions ($\omega=1$). The solid line has been obtained
    from the numerical Legendre Transform of $\mu(\lambda)$ from
    Eq. (\ref{mu_noa2}). The dashed line corresponds to the
    uncorrelated result obtained in Eq. (\ref{piuncorr}), while the
    dot-dashed line is the large deviation function of the Poisson
    distribution of average 1. For completeness, we also show the
    large deviation function of the Poisson distribution with average
    $1/\sqrt{2}$ (dotted line).  }
\end{figure}

In order to get an expression of $\pi(n)$ one should compute the
Legendre Transform of $\mu(\lambda)$. This seems not feasible
analytically, but can be achieved numerically. The function $\pi(n)$
is plotted in Fig. \ref{pi_noa2}, together with the result of
uncorrelated estimation of Eq. (\ref{piuncorr}), as well as two
Poisson distributions, of average $\omega$ and $\omega/\sqrt{2}$.  One
can see that the two estimations carried out are in very close
agreement, apart from a slight difference in the behavior of the right
tail. Moreover, the two new results presented here are clearly
different from the Poisson distribution with mean $\omega$ [denoted
Poisson(1) since $\omega=1$ in Fig. (\ref{pi_noa2})], which is
narrower, and underestimates extreme events, characterized by either
very few or many collisions with respect to typical realizations.

\subsubsection{Sonine perturbation and cumulants}
We now further exploit the property that for $\lambda=0$ the solution
of the Boltzmann equation (\ref{eq:eigenvalue}) is exactly a
Gaussian. This incites us, for $\lambda$ close to 0, to search for a
solution $\tf(\bv, \lambda)$ as a small perturbation of a Gaussian
distribution. One of the most useful expansions in kinetic theory is
the Sonine polynomials expansion \footnote{For a general treatment of
Sonine polynomials in kinetic theory, see {\it e.g.}  \cite{chapman60}
and references therein.}.  In practice this expansion consists in
looking for solutions expressed as a Gaussian times a series of Sonine
polynomials, denoted $S_n(x)$:
\begin{equation}
  \label{series}
\tf(\bv, \lambda)= (2 \pi)^{-d/2} \exp \left(-\frac{v^2}{2 T}\right)
\sum_{n=0}^{\infty} a_n S_n  \left(-\frac{v^2}{2 T}\right)\,\,.
\end{equation}
The first Sonine polynomials are:
\begin{subequations}
\begin{equation}
S_0(x)=1\,\,,
\end{equation}
\begin{equation}
S_1(x)=-x+\frac{d}{2}\,\,,
\end{equation}
\begin{equation}
S_2(x)=\frac{1}{2} x^2 - \frac{d+2}{2} x + \frac{d (d+2)}{8}\,\,.
\end{equation}
\end{subequations}
These polynomials have the property of being orthogonal with respect
to a Gaussian measure in dimension $d$, and are therefore related to
Laguerre polynomials. From this feature, it follows that the
coefficient of the series (\ref{series}) $a_0=1$ and that
$a_1=0$. Hence, the first nontrivial correction to the Gaussian
approximation comes from the term proportional to $a_2$ in the
expansion (\ref{series}).  The procedure to get an estimate of the
coefficient $a_2$ consists of solving a closed system of equations
obtained projecting the equation (\ref{eq:eigenvalue}) onto the first
velocity moments :
\begin{equation}
\mu(\lambda) m_n = \nu_n\,\,,
\label{eq:momentsbis}
\end{equation}
where:
\begin{equation}
\label{eq:moments}
m_n(\lambda)=\int \dd \bv \tf(\bv, \lambda)\,\,,
\end{equation}
and $\nu_n(\lambda)$ denotes the collisional moment of order $n$
defined in Eq. (\ref{collmom}). Truncating the expansions
(\ref{series}) up to the second Sonine polynomial, one gets for the
moments $m_n$:
\begin{subequations}
\begin{equation}
m_0=1\,\,,\qquad m_2=d T(\lambda)\,\,,
\end{equation}
\begin{equation}
m_4=d(d+2) T^2(\lambda) \,\
(1+a_2(\lambda))\,\,.
\end{equation}
\label{eq:newso}
\end{subequations}
The expression of the first collisional moments in the Sonine
approximation is given in the Appendix \ref{app:coll}. Hence, taking
the moment equation (\ref{eq:moments}) for $n=0$, 2 and 4 gives a
closed system of equations for $\mu(\lambda)$, $T(\lambda)$ and
$a_2(\lambda)$. This system can be solved perturbatively expanding its
solutions in power series around $\lambda=0$. The cumulants of $\cN$
obtained by means of the latter expansion turn out to be remarkably
accurate. The expansion of $\mu(\lambda)$ up to the 3$^{\textrm{rd}}$
order in $\lambda$ gives access to the first three cumulants, which
read:
\begin{subequations}
  \begin{align}
  \frac{\langle \cN \rangle_c}{\omega t} & = 1 \,\,, \\ \frac{\langle
  \cN^2 \rangle_c}{\omega t} & =\frac{9}{64} \left(8+\frac{1}{4
  d+3}\right)\,\,,\\ \frac{\langle \cN^3 \rangle_c}{\omega t} &
  =\frac{28 d (64 d (320 d+729)+35775)+257391}{8192 (4 d+3)^3}\,\,,
\end{align}
\end{subequations}
As can be noted from these values, the variance of the number of
collisions already deviates by more than $12.5\%$ from its Poisson
value (equal to unity).

\section{Numerical Results}
\label{sec:simulation}
In this section, we compare the theoretical results obtained in the
previous sections against numerical simulations. In addition to the
two aforementioned techniques of Molecular Dynamics and Direct
Simulations Monte Carlo (see Fig. \ref{fig:fft}), we have shaped a
third numerical tool, constructing a Monte Carlo algorithm in order to
directly solve the eigenvalue equation (\ref{eq:eigenvalue}). We start
by setting the stage for the latter method, before briefly commenting
on the Molecular Dynamics simulations used to measure the statistics
of the number of collisions suffered by tagged particles.

\subsection{Monte Carlo approach}
\begin{figure}
\includegraphics[width=0.46\textwidth,clip=true]{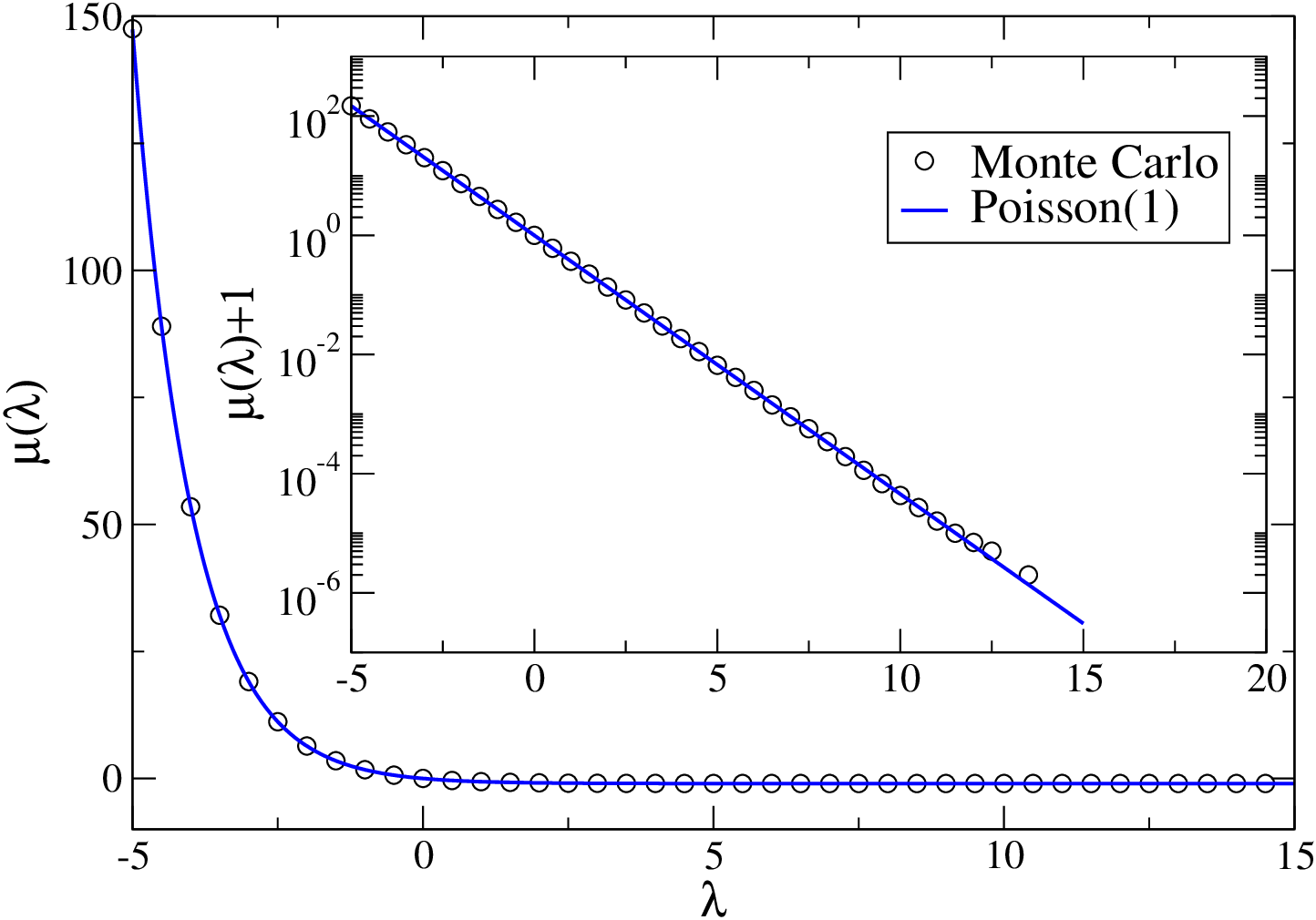}
\caption{\label{fig:maxwell} (Color online) The circles are the
  numerical measurements of $\mu (\lambda)$ for the Maxwell model. The
  solid line is the generating function of the Poisson
  distribution. The inset shows the same data, shifted vertically by
  one (in such a way that $\mu(\lambda) + 1$ is always positive), in
  semi-logarithmic scale.}
\end{figure}

\begin{figure}
 \includegraphics[clip=true,width=0.46 \textwidth]{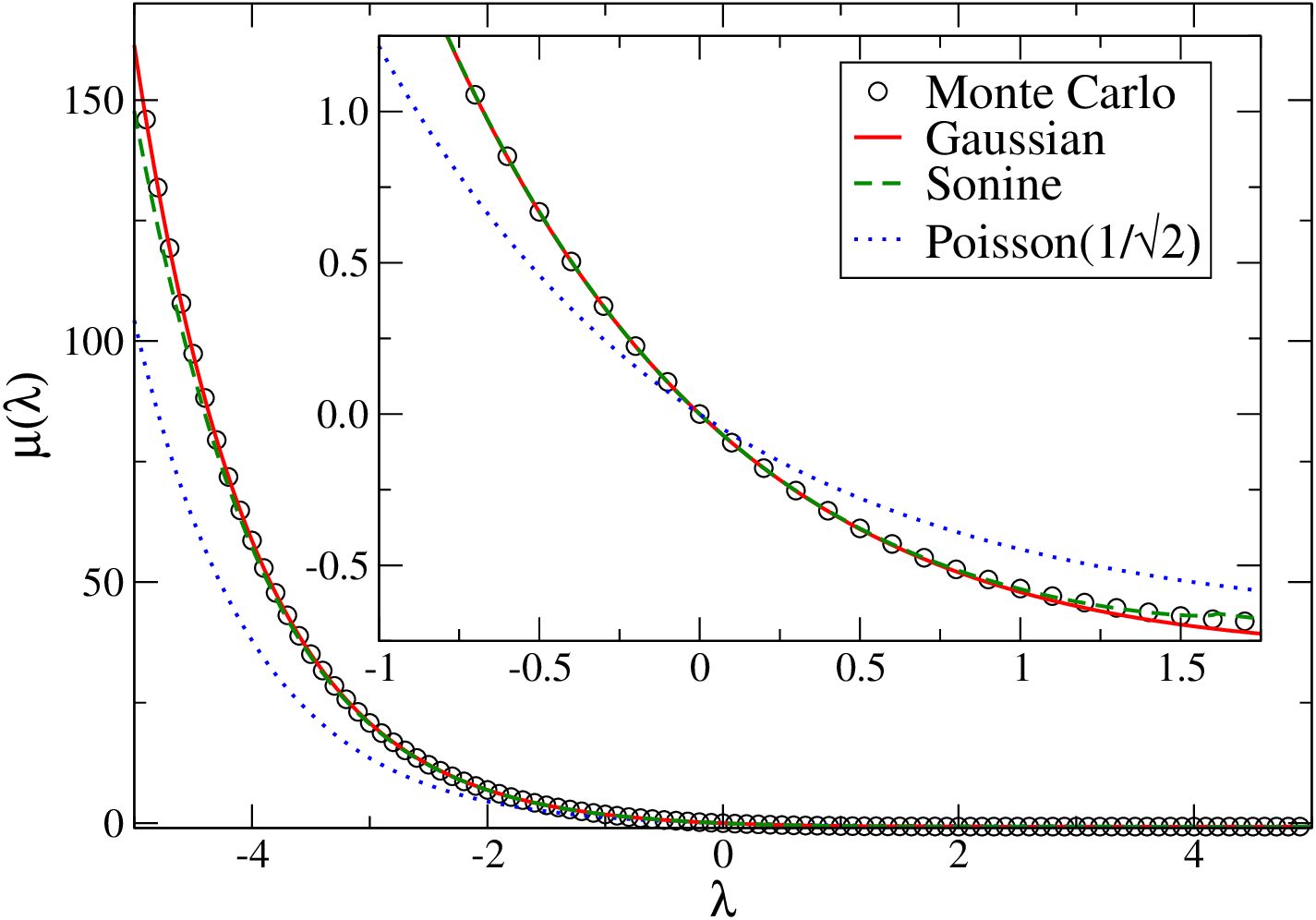}
 \caption{\label{fig:mua2}(Color online) Numerical measurement of
   $\mu(\lambda)$ for the hard-sphere model. The circles show the
   results of Monte Carlo simulations. The solid line is the
   theoretical prediction within the Gaussian approximation. The
   dashed line is the theoretical prediction (obtained solving
   numerically (\ref{eq:momentsbis}-\ref{eq:moments}) in the framework
   of the first Sonine correction).  Finally the dotted line is the
   generating function of a Poisson distribution of parameter
   $\omega/\sqrt{2}$, which should asymptotically dominate for large
   $\lambda$ (see Fig. \ref{fig:mu_largelambda}). The inset shows a
   zoom near $\lambda=0$, where the difference between Gaussian and
   Sonine orders can only be appreciated for the larger values of
   $\lambda$ displayed.}
\end{figure}
\hfill
\begin{figure}
  \includegraphics[clip=true,width=0.46
  \textwidth]{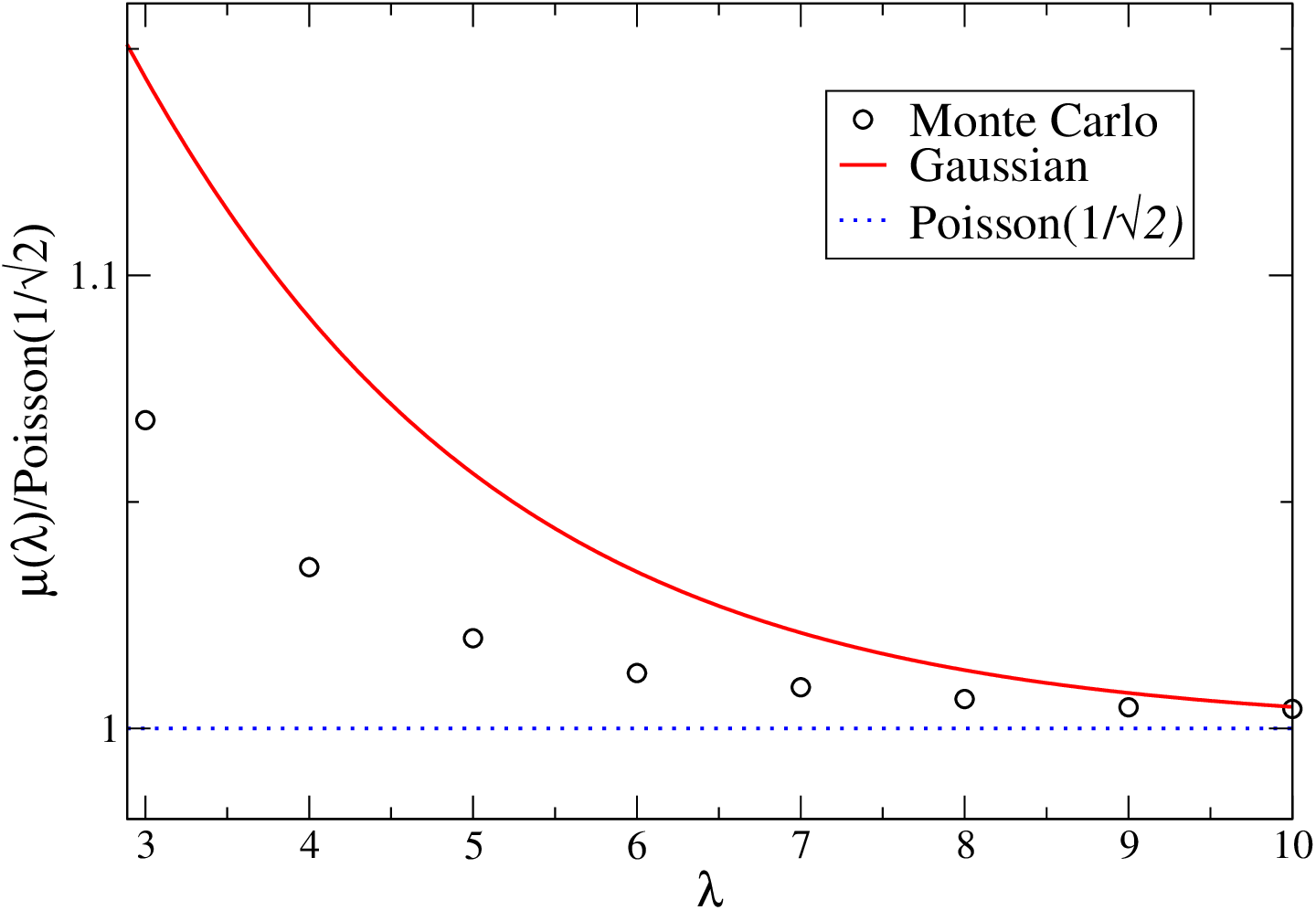}
  \caption{\label{fig:mu_largelambda} (Color online) Same data as in
    Fig. \ref{fig:mua2}, divided by the generating function of a
    Poisson distribution of parameter $\omega/\sqrt{2}$.}
\end{figure}

\begin{figure}
 \includegraphics[clip=true,width=0.46 \textwidth]{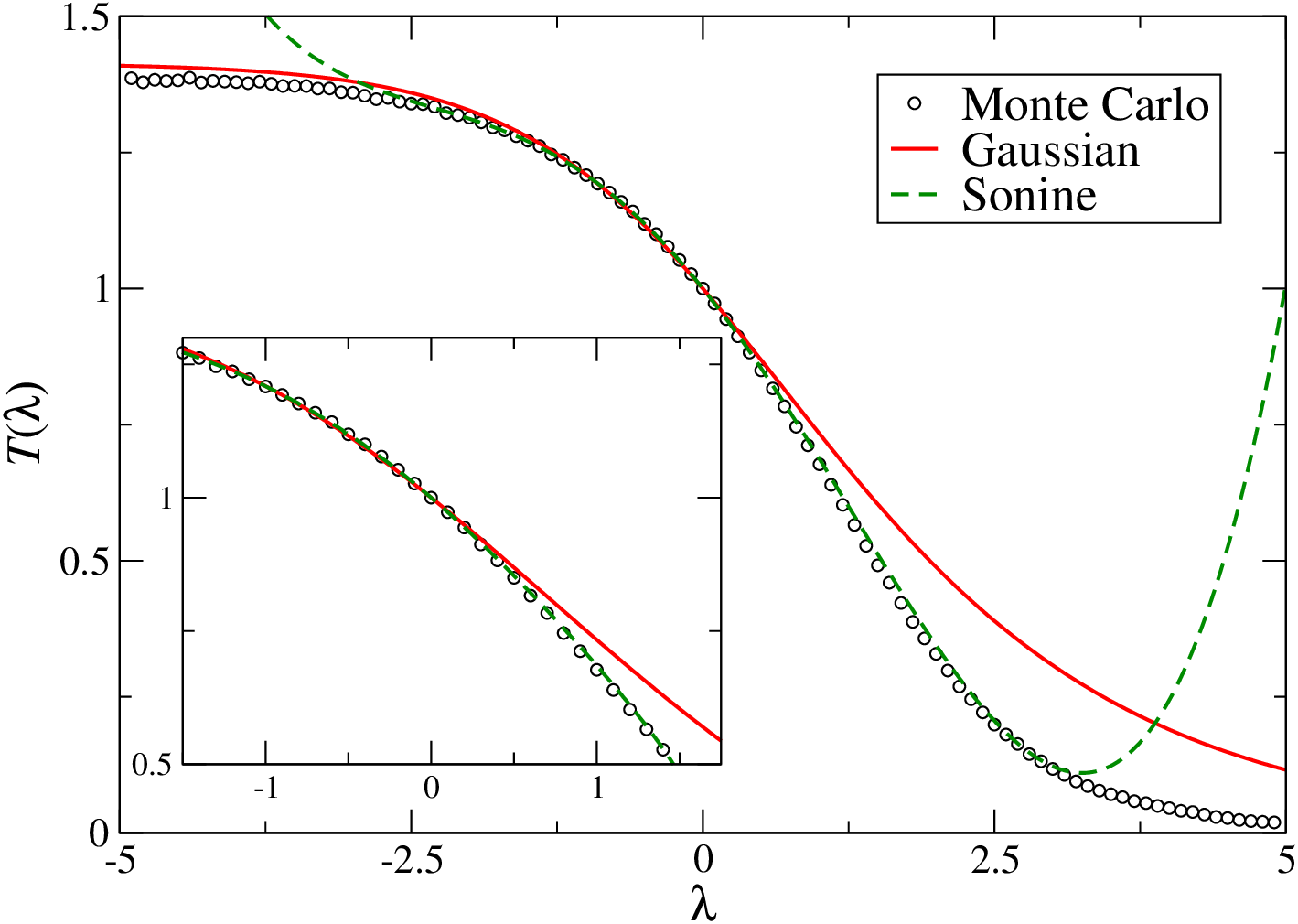}
\caption{\label{fig:temp} (Color online) Temperature (defined as the
  variance) of the eigenfunction $\tf (\bv, \lambda)$ measured in
  Monte Carlo simulations. The solid and dashed lines are the
  predictions of the Gaussian and Sonine approximations respectively.}
\end{figure}

\begin{figure}
  \includegraphics[clip=true,width=0.46
  \textwidth]{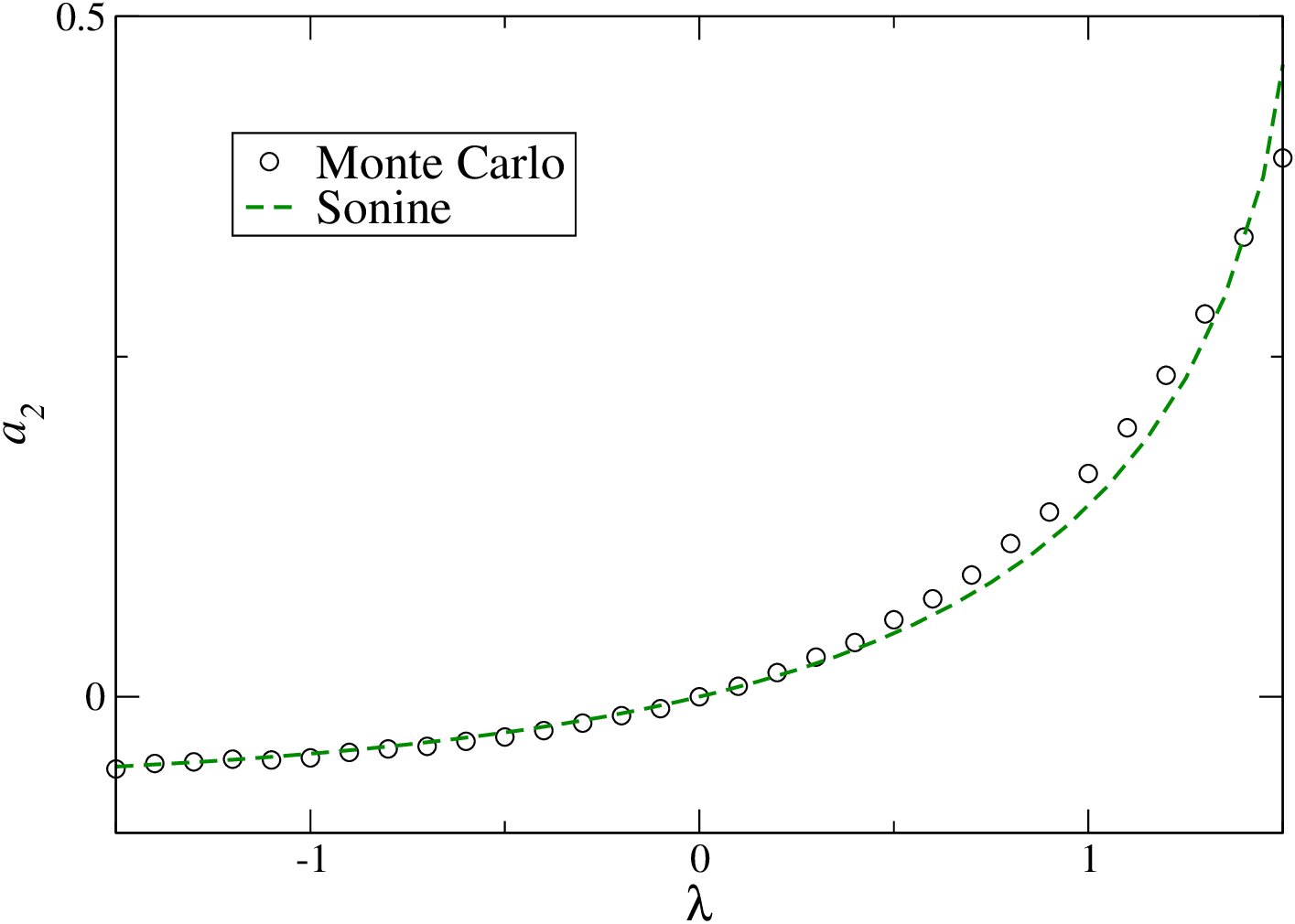}
  \caption{\label{fig:a2} (Color online) Coefficient $a_2$
  ($\lambda$-dependent velocity kurtosis), measured in a Monte Carlo
  simulation. The dashed line corresponds to the numerical solution of
  the system (\ref{eq:newso}), which is correct only for small values
  of $a_2$. }
\end{figure}
In order to derive an algorithm for solving Eq. (\ref{eq:eigenvalue}),
it is useful to rewrite this equation in the form:
\begin{equation}
\label{probann}
\mu(\lambda) \tf(\bv, \lambda)= \ee^{-\lambda} I[\tf|\phi] + 
(1-\ee^{-\lambda}) A[\tf|\phi]\,\,,
\end{equation}
where 
\begin{equation}
I[f|g]=\int \dd \bv \int' \dd \bs (\bv_{12}\cdot \bs) [f_1^{**}
g_2^{**} -f_1 g_2]\,\,,
\end{equation}
is the collision integral describing the elastic collision between two
particles having respectively a velocity pdf $f$ and $g$. We are using
the short-hand notation $f_i^{**} \equiv f(\bv^{**}_i)$. The
functional $A[f|g]$ is the loss term of the above collision
integral. It actually describes the statistics of hard spheres which
annihilate after each collision:
\begin{equation}
A[f|g]=-\int \dd \bv \int' \dd \bs (\bv_{12}\cdot \bs) f_1 g_2\,\,.
\end{equation}
In the context of the nonlinear Boltzmann Equation, i.e. when $\phi =
\tf$ in Eq. (\ref{probann}), and for positive values of $\lambda$, the
rhs of Eq. (\ref{probann}) exactly describes the time evolution of the
velocity pdf of the probabilistic ballistic annihilation process. This
model, introduced in \cite{coppex04}, consists in a system of $N$
particles which move ballistically, and interact when at contact. The
interaction may be an elastic collision, with probability
$\ee^{-\lambda}$, or an annihilation (with probability
$1-\ee^{-\lambda}$). In the context of the linear Boltzmann equation,
this process can be extended in the following way. Consider a set of
$N$ independent systems, one of each is just a single particle,
characterized by its velocity $\bv_i(t)$ ($i=1,\dots,N$), assuming
spatial homogeneity. If each of these systems (particles) evolves in a
hard sphere gas (in the thermodynamic limit) thermalized at
temperature $T_0$, with a one point velocity pdf $\phi$, and can both
collide elastically (with probability $\ee^{-\lambda}$), or annihilate
(with probability $1-\ee^{-\lambda}$), then, the reduced one point
velocity pdf $g(\bv,t)= \langle \Sigma_i^N \delta(\bv-\bv_i(t))
\rangle$ will verify the Boltzmann Equation
\begin{equation}
\label{blabla}
\frac{\partial g(\bv,t)}{\partial t} = L_{\lambda} g(\bv, t)\,\,.
\end{equation}
Since $N(t)=\int \dd \bv g(\bv,t)$ is the number of particles at time
$t$, and since we know that for long times $g(\bv, t) \sim \ee^{\mu
t}$, it is clear that for long times one has that $N(t) \sim N_0
\ee^{\mu t}$. Note that the present interpretation implicitly assumes
positive values of $\lambda$. Moreover, the particles can only
annihilate or collide; then the total number of particles can only
decrease, and hence $\mu$ must be negative. The above observations
provide a numerical tool for measuring $\mu(\lambda)$ as the decay
rate of the total number of particles. The main difficulty, which we
have successfully addressed, with the above algorithm, is that the
number of particles constantly decreases, and there is no steady state
but the trivial state $N=0$. To circumvent this practical difficulty
(that would lead to somewhat noisy statistics in the simulations), we
have introduced an external source of particles (systems), acting in
such a way that the total number of particles is conserved. If every
time that an annihilation takes place, a new particle is inserted as
the clone of one of the $N-1$ remaining particles (chosen uniformly
among this population), then the evolution equation of $g(\bv,t)$
reads:
\begin{equation}
\partial_t g(\bv,t)=L_{\lambda} g(\bv, t)+ s  g(\bv, t)\,\,,
\end{equation}
where $s$ is a constant rate. At late times, if the above equation has
a steady state, the largest eigenvalue of the operator $L_{\lambda} +
s$ vanishes, and hence one has that $\mu(\lambda) + s =0$.  Finally
one can measure $\mu(\lambda)$ simply as (minus) the steady state
average of the number of particles injected by the external source.

So far, we showed how to construct a Markov chain in order to simulate
the eigenvalue equation (\ref{eq:eigenvalue}) for positive values of
$\lambda$. For negative values of $\lambda$ the procedure is almost
identical. Introducing a new time scale ${\tilde t}= (2 \ee^{-\lambda}
-1)t$, Eq. (\ref{blabla}) can be rewritten as:
\begin{equation}
  \partial_{\tilde t} g(\bv,{\tilde t})= \frac{\ee^{-\lambda}}{2
    \ee^{-\lambda}-1} I[g|\phi]-
\frac{\ee^{-\lambda}-1 }{2 \ee^{-\lambda}-1} A[g|\phi]\,\,.
\end{equation}
All the previous physical interpretations and remarks still hold,
except that now instead of having annihilation (with probability
$(\ee^{-\lambda}-1)/(2 \ee^{-\lambda}-1)$), one has duplication (or
cloning). Hence $\mu(\lambda)$ will be positive, and one can add an
external source in order to remove particles when new particles are
created.  Summarizing, the algorithm proceeds as follows:
\begin{enumerate}
\item[(o)] The velocity of the $N$ particles are stored in a $N \times
  d$ matrix.  A scalar $s$ is set to 0. A (small) time step
  $\dd t$ is chosen.
\item[(i)] A particle is chosen randomly in the population 
with uniform probability.
Its velocity is denoted $\bv_1$.
\item[(ii)] An interaction is accepted with a probability
  $\propto \bv_{12} \cdot \bs \Theta (\bv_{12} \cdot \bs)\times \dd
  t$, where $\bs$ is a random direction in $d$ dimensions, and $\bv_2$
  a $d$-dimensional zero-mean Gaussian random variable of variance
  $T_0$. In practice, $\dd t$ has to be chosen in such a way that
  $|\bv_{12} \cdot \bs| \dd t$ is always smaller than (or equal to)
  one.
\item[(iii)] When the interaction is accepted, if $\lambda>0$
  (resp. $\lambda<0$), the particle will have a post-collisional
  velocity $\bv_1^{*}$ with probability $\ee^{-\lambda}$
  $\left(\hbox{resp. } \frac{\ee^{-\lambda}}{2 \ee^{-\lambda}-1}
  \right)$) or will be removed (resp. duplicated) otherwise.
\item[(iv)] If the particle has been removed (resp. duplicated) in
  step (iii), one of the $N-1$ remaining particles is chosen randomly
  and uniformly, and is duplicated (resp. removed). $s$ is increased
  by 1.
\item[(v)] Time is increased by an amount $\dd t$ (resp. $\dd {\tilde
t}$).
\item[(vi)] Back to (i).
\end{enumerate}
This algorithm bears some similarities with the approach proposed in
\cite{giardina}, also intended to directly measure large deviation
functions. Nonetheless, the version proposed here is more inspired by
some variants of the DSMC algorithm for systems which do not conserve
the total number of particles \cite{et2002,PTD02,TK03}.  In order to
check the reliability of the algorithm, we have first performed
simulations in the case of Maxwell molecules, where the number of
collisions is exactly distributed following the Poisson distribution
(cf. the Introduction). The measurements of $\mu(\lambda)$ for this
particular model are shown in Fig.  \ref{fig:maxwell}, together with
the generating function of the Poisson distribution. The agreement is
excellent.  In the simulations the temperature scale is set by the
temperature of the heat bath $T_0$, which we set to unity. The time
scale is set by the mean free time, which we also set to unity.  In
the simulation data, the mean collision frequency is therefore
$\omega=1$.

In the case of the hard-sphere model, a measurement of the largest
eigenvalue $\mu(\lambda)$ is shown in Fig. \ref{fig:mua2}. One can see
that the numerical results are in very good agreement both with the
Gaussian and the Sonine approximations. Figure
\ref{fig:mu_largelambda} shows the large $\lambda$ behavior of
$\mu(\lambda)$ divided by the prediction (\ref{mu_largelambda}).  In
this case, a good agreement between the numerical data and the
theoretical predictions is found. In the framework of the above
described Monte Carlo algorithm, it is also possible to measure the
stationary velocity pdf $g(\bv, t=\infty,\lambda)$ which is equal to
the eigenfunction $\tf(\bv, \lambda)$ associated with $\mu(\lambda)$.
Hence one can also compare the analytical predictions for the
temperature $T(\lambda)$ and for the Sonine coefficient $a_2
(\lambda)$ with the Monte Carlo results, see Figs. \ref{fig:temp} and
\ref{fig:a2}.
In Fig. \ref{fig:temp}, one can see the temperature $T(\lambda)$. Here
the Gaussian approximation is already able to capture the general
behavior of this ``effective'' temperature, and for small values of
$\lambda$, the Sonine corrections compare very well with the
simulation data.  In Fig. \ref{fig:a2}, one can see the coefficient
$a_2$ as a function of $\lambda$. As expected, for small values of
$\lambda$ the expansion carried out in the previous section correctly
describes the result of the simulations.

\subsection{Molecular Dynamics}
\begin{figure}
\includegraphics[width=0.46\textwidth,clip=true]{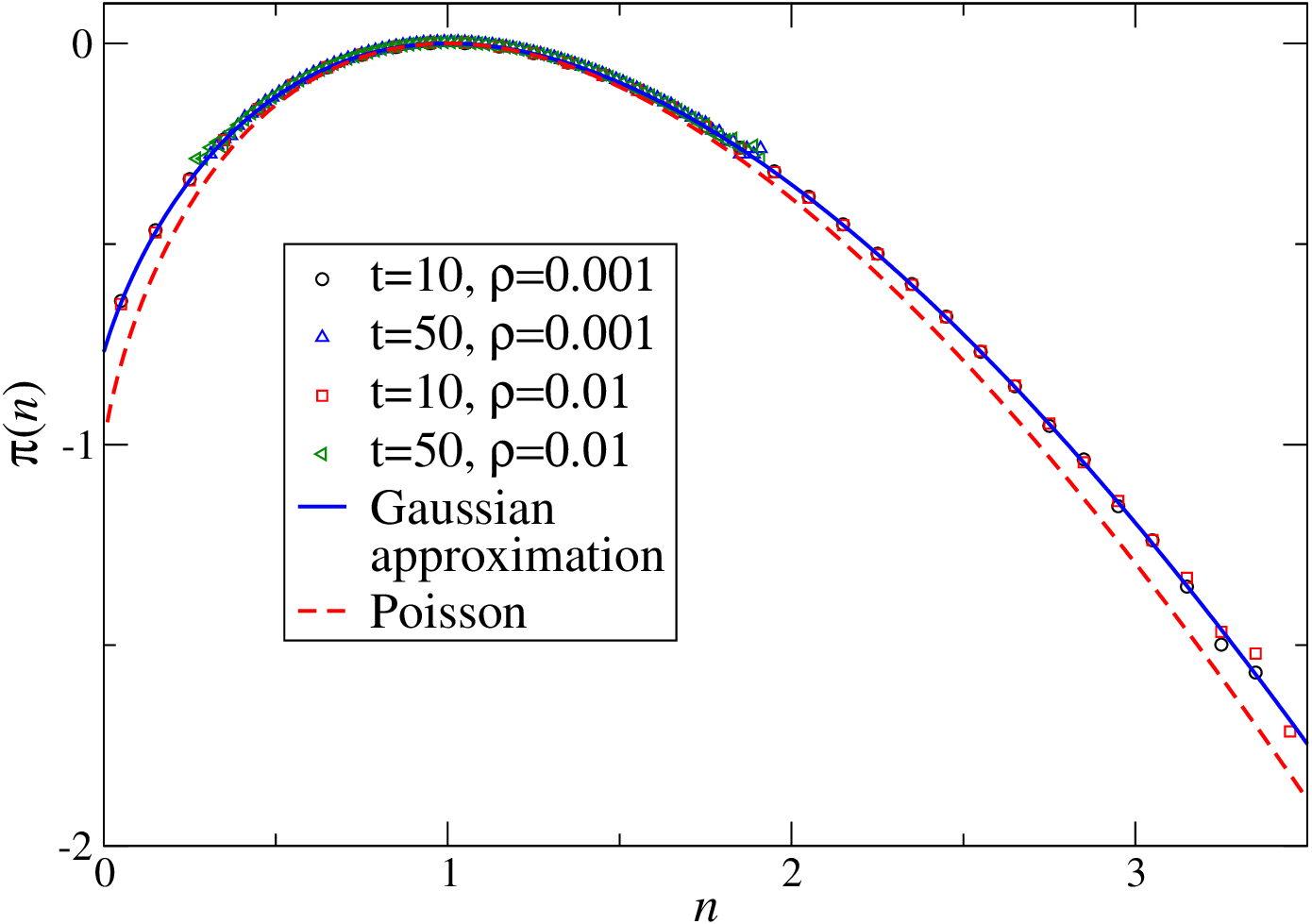}
\caption{\label{fig:pi_md} (Color online) Large deviation function
  $\pi(n)$ of the number of collisions $\cN$ suffered by each single
  particle. Symbols show the results of Molecular Dynamics
  simulations.  The solid line is the result in the framework of the
  Gaussian approximation, and the dashed line is the large deviation
  function of the Poisson distribution, of average 1: $\pi(n) = n
  -n\log n-1$.}
\end{figure}

The most direct way to measure the large deviation function of the
number of collisions $\cN$ is of course to count it in a (numerical)
experiment, and then construct its probability distribution
function. To this end, we have performed Molecular Dynamics (MD)
simulations of a two dimensional hard-disk gas of $N=10^3$ particles
of diameter $\sigma=1$ at density $\rho=N/V=10^{-3} \sigma^{-2}$ and
$10^{-2} \sigma^{-2}$. The particles evolve in a square box with
periodic boundary conditions, and the time is measured in mean free
time units, in such a way that $\langle \cN \rangle = t$. We measured
the statistics of the number of collisions $\cN$ suffered by each
particle. The value of the first cumulants is reported in Table
\ref{tab:cumulants}, together with the Poisson prediction, as well as
the results from the Gaussian and Sonine approximations. It seems
that, when time increases, the cumulants converge towards a value
which is close to the Sonine predictions. However, it must be noted
that even if increasing time makes finite time corrections smaller,
the statistics become poorer and poorer. The agreement with the
results from the Sonine approximation is very good, while the Gaussian
order already provides a reliable estimation.  This is further shown
in Figure \ref{fig:pi_md}, where two measurements of the large
deviation function of the number of collisions for two different
times, $t=10$ and $t=50$ (with $\sigma=1$ and $T_0=1$), are compared
to the numerical inverse Legendre Transform of $\mu(\lambda)$ at
Gaussian order (given by Eq.  (\ref{mu_noa2})). Here again, the
Poisson prediction is distinctly off.

\begin{table}
\caption{\label{tab:cumulants} Cumulants for the number of collisions
    $\cN$ from MD simulations, and comparison with the Gaussian and
    Sonine approximations (time is measured in units of the mean-free
    time). }
\begin{ruledtabular}
\begin{tabular}{lcccr}
&$\langle \cN \rangle_c /t$ & $\langle \cN^2 \rangle_c /t$ & 
$\langle \cN^3 \rangle_c /t$\\
\hline
$t=10$  & 1. & 1.1228 & 1.1282 \\
$t=50$  & 1. & 1.1354 & 1.1045   \\
Poisson & 1  & 1       & 1       \\
Gaussian& 1  & 1.125   & 1.1289 \\
Sonine  & 1  & 1.1377  & 1.1073  \\
\end{tabular}
\end{ruledtabular}
\end{table}

\section{Conclusion}
We have shown that the collisional statistics of the hard sphere gas
exhibits clear deviations from the Poisson distribution. These
deviations have been consistently quantified both analytically and
numerically.  In the analytical treatment, the cumulant generating
function $\mu(\lambda)$ plays a pivotal role. The eigenvalue equation
defining this object can be fruitfully interpreted in terms of
population dynamics with annihilation and cloning events, which
eventually leads to an efficient algorithm allowing to compute the
various quantities involved in the theoretical analysis. The
corresponding numerical method, of Monte Carlo type, should not be
confused with the more conventional Direct Simulation Monte Carlo
technique, which we also implemented, and that is intended to solve a
different kinetic equation (the Boltzmann equation). Finally, a third
numerical method (Molecular Dynamics) was employed. These three routes
provide complementary and valuable results, that strongly support our
predictions.

Interestingly, the present formalism can be extended to the study of
out of equilibrium systems, such as granular gases, where the question
of the collisional statistics has been the focus of recent interest
\cite{Luding,Pago,Blair,Paolotti,Falconetal}. In particular, the
strong effect of dissipation on the distribution of free flight times
reported in \cite{Blair,Paolotti} calls for further investigations.
In addition, if one considers a gas of inelastic smooth hard spheres,
kept in a steady state by a velocity independent force (as e.g. a
vibrating wall of the container, or a stochastic force acting
independently on each particle), then the phase space volume of the
system has been reduced, after a time $t$, by a factor
$(1-\alpha)^{\cN(t)}$, where $\alpha$ is the coefficient of normal
restitution \cite{poschel01}. Hence one sees that in this
non-equilibrium system, the number of collisions can be exactly
identified, up to a constant prefactor, with the integrated phase
space contraction rate. This quantity has already been the subject of
many works in non-equilibrium statistical mechanics, and is often
intimately related with the irreversible entropy production (see
e.g. \cite{EPLPaolo} and references therein).

\begin{acknowledgments}
  The authors acknowledge useful discussions with J. Piasecki, J.
  M. J. van Leeuwen, M. H. Ernst, D. Frenkel, H. van Beijeren and
  P. Krapivsky.  This work was supported by the French Ministry of
  Education through a ANR-05-JCJC-44482 grant.
\end{acknowledgments}

\appendix
\section{Comment on the free flight time distribution}
\label{app:fftwrong}
We shall give here some additional arguments on the incorrectness
of Eq. (\ref{eq:fft}), when used with a Gaussian weight $\phi$,
instead of the weight $\phi_{coll}$ defined in
Eq.(\ref{eq:phicoll}). We shall refer to this (erroneous) distribution as
$P^W_{FFT}$:
\begin{equation}
\label{eq:pfftwrong}
P^W_{FFT}(t)=\int \dd \bv \phi(\bv) P(t|\bv)\,\,.
\end{equation}
A first argument bears on the inconsistency between the above relation
and the definition of the collision frequency $\omega=\langle r(v)
\rangle$, where the brackets denote an average over a Gaussian weight.
Indeed, the average time $\tau$ between two subsequent collisions
(mean free time) of a given particle is equal to the inverse of the
collision frequency: $\tau=1/\langle r(v)\rangle$. Besides, from
Eq. (\ref{eq:pfftwrong}) the mean free time $\tau$ is obtained as:
\begin{equation}
\tau=\int_0^{\infty}\dd t \,\,t P^W_{FFT}(t)=
\left\langle \frac{1}{r(v)} \right \rangle
\ne \frac{1}{\langle r(v) \rangle}.
\label{eq:a2}
\end{equation}
Hence we see that expression (\ref{eq:pfftwrong}) is in contradiction
with the definition of the collision frequency. On the other hand, the
counterpart of Eq. (\ref{eq:a2}) with the distribution provided by
(\ref{eq:fft}) and (\ref{eq:phicoll}), yields
\begin{equation}
\tau = \int_0^\infty dt\,\, t \int \dd \bv \,\frac{r^2(v)}{\omega} \, 
e^{-r(v) t}\, \phi(\bv) \, =\, \frac{1}{\langle r
\rangle},
\end{equation}
which is the required relation.

Second, expression (\ref{eq:pfftwrong}) explicitly slightly differs
from the correct free flight time distribution in two limiting cases.
\begin{itemize}
\item At small values of $t$, $P^W_{FFT}$ behaves as:
\begin{equation}
P^W_{FFT}(t) \sim \langle r(v) \rangle - \langle r(v)^2 \rangle t +
\cO(t^2)\,\,,
\end{equation}
while the true distribution behaves as
\begin{equation}
  P_{FFT}(t) \sim \frac{\langle r(v)^2 \rangle}{\omega} - 
\frac{\langle r(v)^3 \rangle}{\omega^2} t + \cO(t^2)\,\,,
\end{equation}
so that even the value in $t=0$ is different, although this difference
is numerically small (see also Fig. \ref{fig:fftwrong}). For instance,
in $d=2$ one has:
\begin{equation}
\frac{\langle r(v)^2 \rangle}{\omega^2}\simeq 1.06354
\end{equation}
and
\begin{equation}
\frac{\langle r(v)^3 \rangle}{\omega \langle r(v)^2 \rangle}
\simeq 1.13962\,\,.
\end{equation}
\item The large time behavior of the probability also slightly differs
  from Eq. (\ref{eq:fftlt}):
\begin{equation}
\label{eq:fftltw}
P_{FFT}^W \sim \exp\left(-\frac{\omega t}{\sqrt{2}} \right)
2^{\frac{d-1}{2}} \omega \left(2-\frac{2}{d}+ \frac{\sqrt{2} \omega
t}{d} \right)^{-d/2}\,\,.
\end{equation}
In the above expression, the leading exponential term remains the same
as in Eq. (\ref{eq:fftlt}), since it is only determined by the minimum
of the function $r(v)$. Nonetheless, the subleading prefactor is
slightly different, as it can be appreciated in
Fig. \ref{fig:fftwrong}.

\begin{figure}
  \includegraphics[width=0.46\textwidth,clip=true]{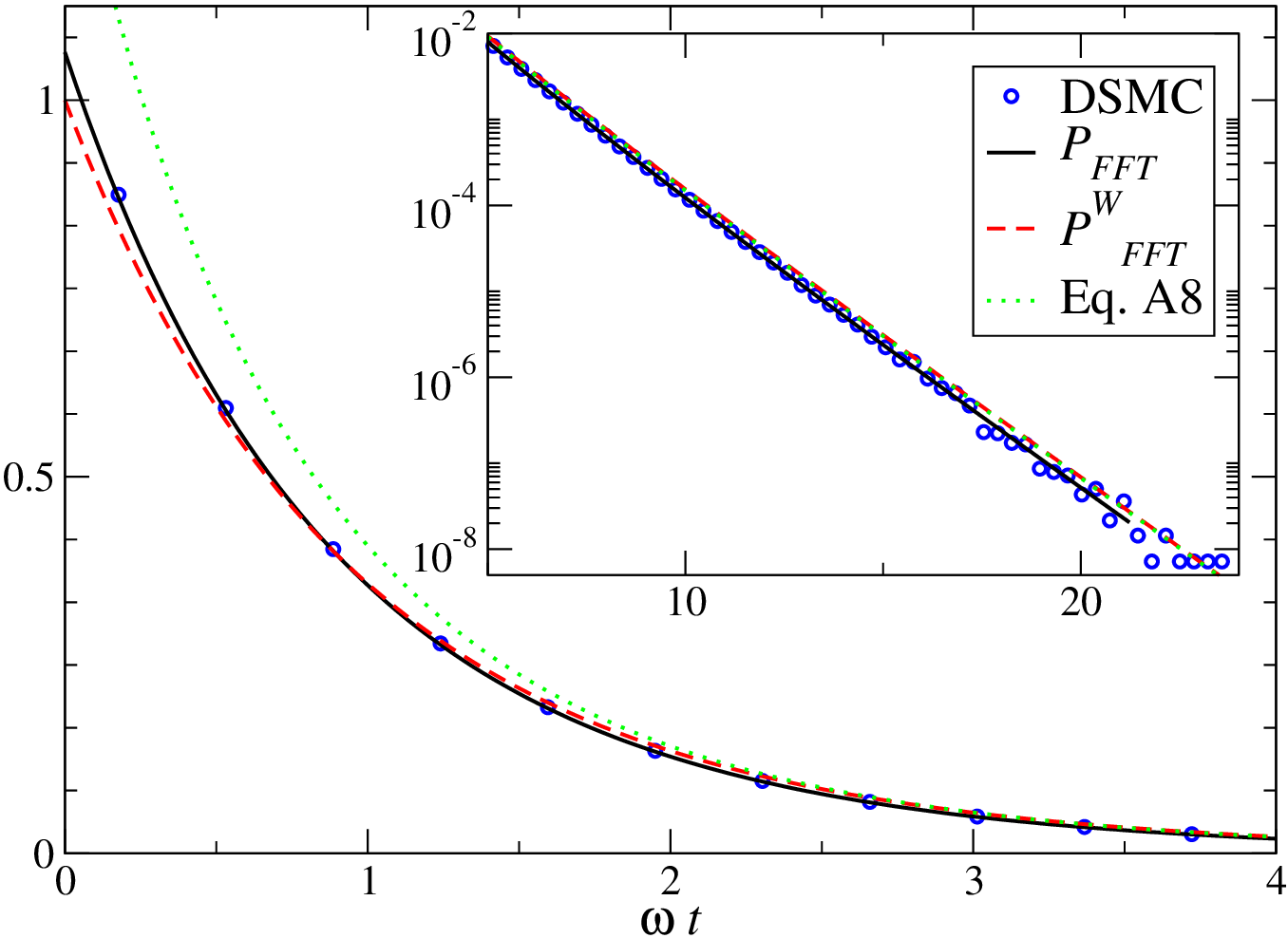}
  \caption{\label{fig:fftwrong} (Color online) Free flights time
    distribution of a hard disc gas. The circles corresponds to the
    results of Molecular Dynamics (MD) simulations at density
    $\rho=0.01 \sigma^{-2}$. The full line is the numerical
    integration of Eq. (\ref{eq:fft}). The dashed line is the
    numerical integration of $P_{FFT}^W$, and the dotted line its
    large time behavior described by Eq. (\ref{eq:fftltw}). The inset
    shows the same data in semi-logarithmic scale, where it is
    possible to note that at large times the prediction of $P_{FFT}^W$
    is always above the true distribution. Time is measured in units
    of the mean free time ($\omega=1$).}
\end{figure}

\end{itemize}

\section{Free flights time distribution for Very Hard Particles}
\label{app:vhp}
\begin{figure}
  \includegraphics[width=0.46\textwidth,clip=true]{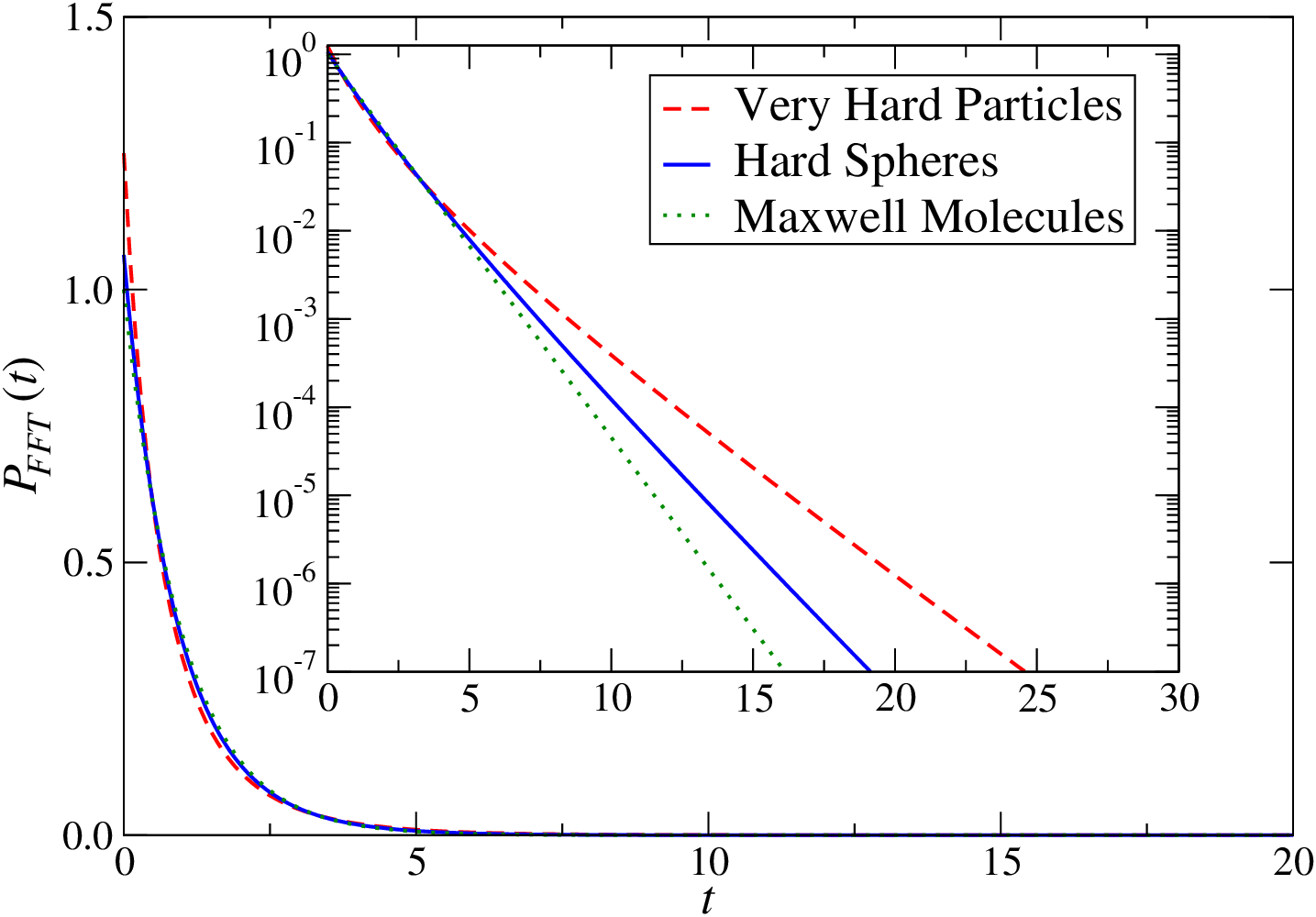}
  \caption{\label{fig:fftvhp} (Color online) Free flights time
    distributions for three different interaction kernels, in two
    dimensions. The (red) dashed line shows the result for VHP
    (Eq. \ref{eq:fftvhp}), the (blue) solid line and the (green)
    dotted line show respectively the results for Hard Spheres and
    Maxwell Molecules. Time is measured in units of the mean free time
    ($\omega=1$).}
\end{figure}

For the sake of completeness, we report in this appendix the result for
the free flight time distribution arising in the very hard particle (VHP)
model. This framework allows for an explicit analytic computation of
$P_{FFT}(t)$. 
The model, introduced by Ernst and Hendriks \cite{ernst79},
consists in a choice of the collision rate proportional to the total
energy of the system. In the case of a single tagged particle, the
corresponding linear Boltzmann equation reads:
\begin{equation}
\label{linboltzvhp}
\frac{\p}{\p t}  f(\bv_1,t)=
\frac{1}{\ell} \int \!\!
\dd \bv_2 \int' \!\!\!\! \dd \bs \  
\frac{(\bv_{12} \cdot
\bs)^2}{\sqrt{T_0}}  \left[f_1^{**} \phi_2^{**} - 
 f_1  \phi_2 \right]\,\,.
\end{equation}
The velocity dependent collision rate is thus quadratic in $v$:
\begin{equation}
\frac{r_{VHP}(\bv)}{\omega_{VHP}}=\frac{1}{2} + \frac{v^2}{2 d
T_0}\,\,,
\end{equation}
where
\begin{equation}
\omega_{VHP} =\frac{2 \pi^{d/2} \sqrt{T_0}}{\ell \Gamma(d/2)}
\end{equation}
is the collision frequency for VHP. Due to the simpler form of the
collision rate, the free flight time distribution can be expressed
analytically:
\begin{multline}
\label{eq:fftvhp}
P_{FFT}^{VHP}(t)= \frac{1}{4} d^{d/2} \ee^{-t/2}  \times \\
   (d+t)^{-\frac{d}{2}-2} \left(4 d^2+(4 t+2) d+t^2\right)\,\,,
\end{multline}
where the time is here in units of the mean collision rate
$\omega_{VHP}$. Note that here the large time behavior is $\sim
\ee^{-\omega t/2}$, which is even slower than the Hard-Sphere gas 
behaviour (which is $\sim \ee^{-\omega t /\sqrt{2}}$) with respect to the
Poissonian case ($\sim \ee^{-\omega t}$). A comparison between these
three distributions is shown in Fig.~\ref{fig:fftvhp}. The observed
large time behaviors suggest that Maxwell and VHP models provide a
upper and lower bounds for the hard-sphere model, a phenomenon reminiscent 
to that observed in
the long time behaviour of dynamics that do not conserve the density
\cite{TK03}.

\section{Saddle-point approximation for the collision rate}
\label{app:saddle}
In this appendix we show how to recover Eq. (\ref{rlarged}) with the
saddle point method. We first note that $r(v)$ can be expressed as the
difference of two different integrals:
\begin{equation}
r(v)=\Omega_{d-1} (I_1(v) - I_2(v))\,\,,
\end{equation}
where
\begin{multline}
I_1(v)=\int_{-\infty}^{+\infty} \dd \vds \int_0^{\pi} \dd \theta \sin
\theta ^{d-2} (v \cos \theta) \times \\ \Theta(v \cos \theta - \vds)
\phi(\vds)\,\,,
\end{multline}
\begin{multline}
I_2(v)=\int_{-\infty}^{+\infty} \dd \vds \int_0^{\pi}\dd \theta \sin
\theta ^{d-2} \vds \times \\ \Theta(v \cos \theta - \vds)
\phi(\vds)\,\,.
\end{multline}
For the first integral it is more convenient to perform first the
integration over the angle $\theta$, which leads to:
\begin{equation}
  I_1(v)=\frac{v}{d-1} \int_{-\infty}^{+\infty} \dd \vds
  \left(1-\left(\frac{\vds}{v}\right)^2\right)^{\frac{d-1}{2}}
  \phi(\vds)\,\,.
\end{equation}
Then, defining the rescaled variables $\vt=v/\sqrt{2 T_0 d}$ and
$\vdst=\vds/\sqrt{2 T_0 d}$ one finds, to leading order in $d$:
\begin{multline}
  I_1(\vt) \simeq \vt \sqrt{\frac{T_0}{\pi}} \int_{-\infty}^{+\infty}
  \dd \vdst \,\, \times \\ \exp \left[d \left(-\vdst^2+\frac{1}{2}
  \ln\left(1-
  \left(\frac{\vdst}{\vt}\right)^2\right)\right)\right]\,\,.
\end{multline}
When $d$ is very large, the above integral is dominated by the maximum
of the function inside the exponential, which turns out to be located
in $\vdst=0$. One can then perform a series expansion around $\vdst=0$
up to the second order. This results in a Gaussian integral which is
easily integrated, and yields:
\begin{equation}
I_1(\tv)\simeq \sqrt{\frac{T_0}{d}} \frac{2 \vt^2}{\sqrt{1+2
\vt^2}}\,\,.
\end{equation} 
As for the second integral involved in the expression of
$r(v)$ the simplest is first to perform the integration over $\vds$;
hence one obtains, to leading order in $d$:
\begin{equation}
I_2(\vt)=-\sqrt{\frac{T_0}{2 \pi}} \int_0^{\pi} \exp\left[ d
    \left(-\vt^2 \cos^2 \theta + \ln \sin \theta \right) \right]\,\,.
\end{equation}
The only maximum of the function in the exponential between $0$ and
$\pi$ is in $\theta=\pi/2$. Then, expanding as usual this function
around the maximum up to second order, and extending the range of
integration from $-\infty$ to $+\infty$, one finds:
\begin{equation}
I_2(\vt)\simeq \sqrt{\frac{T_0}{d}} \frac{1}{\sqrt{1+2 \vt^2}}\,\,.
\end{equation}
Finally, noting that $\Gamma(d/2)/\Gamma((d-1)/2)\stackrel{d \to
  \infty}{\sim} \sqrt{\frac{d}{2}}$, Eq. (\ref{rlarged}) follows.

\section{Free path length distribution for Maxwell molecules}
\label{app:fplmaxwell}
\begin{figure}
  \includegraphics[width=0.46\textwidth,clip=true]{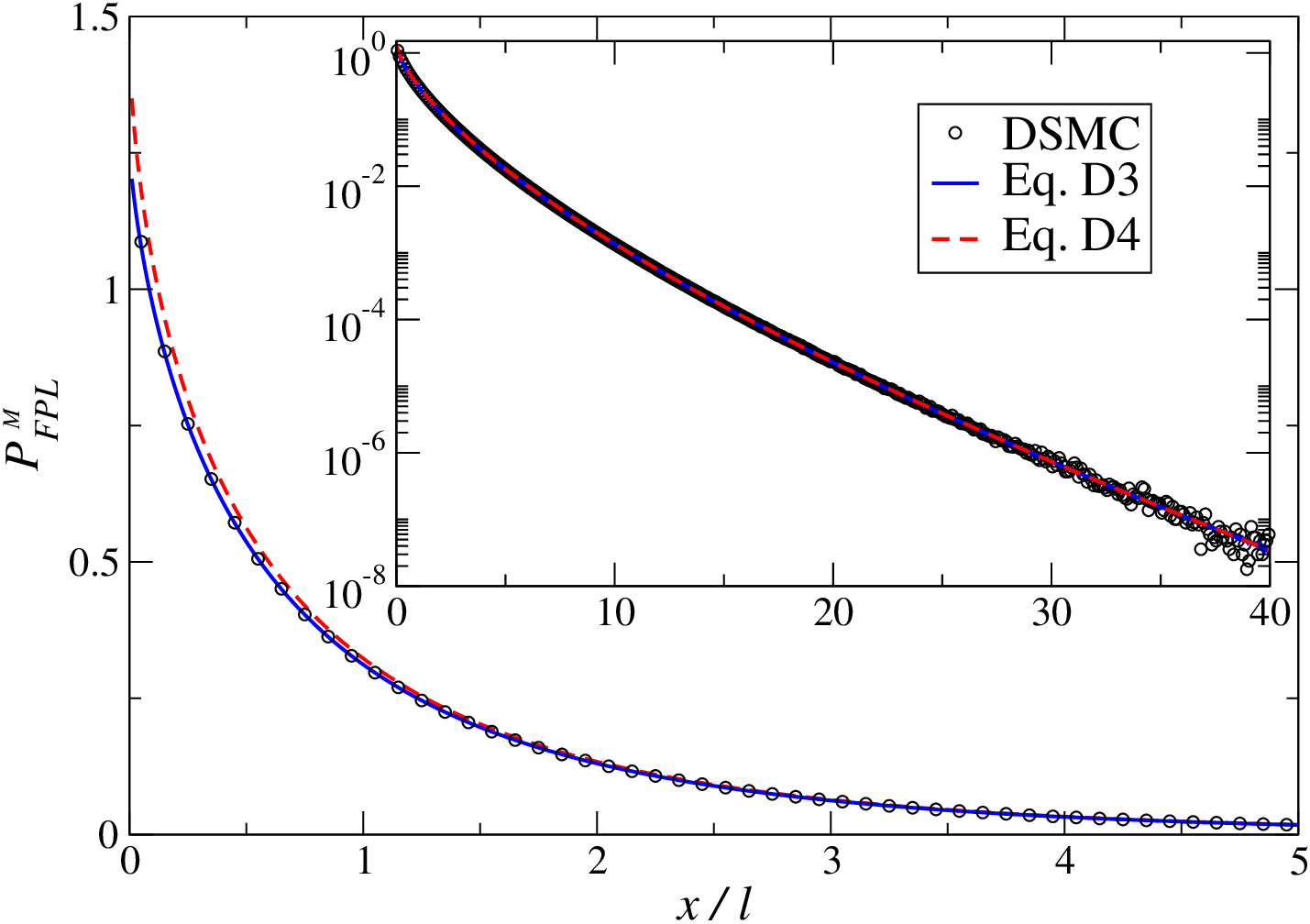}
  \caption{\label{fig:fplmaxwell} (Color online) Free path length
    distribution for Maxwell molecules in two dimensions. The circles
    are the result of a DSMC simulation, while the solid line is the
    numerical integration of Eq. (\ref{eq:fplmaxwell}). The dashed
    line is the large length approximation of
    Eq. (\ref{eq:fplmaxlargex}). All data are reported in units of
    mean free path.}
\end{figure}

In this appendix, we investigate the free path length distribution for
Maxwell molecules. As already mentioned, in this case the collision
rate is a constant, $\omega$, independent of the velocity of the
particle. It follows then that the free flight time probability is
exponential:
\begin{equation}
P_{FFT}^M(t|\bv)=P_{FFT}^M(t)=\omega \ee^{-\omega t}\,\,.
\end{equation}
Besides, the free path length distribution for a given velocity $\bv$
reads:
\begin{equation}
P_{FPL}^M(x|\bv)=\frac{\omega}{v} \exp\left(-\frac{\omega}{v} x
\right)\,\,,
\end{equation}
and the free path length distribution is simply the average of the
above probability over a Gaussian weight (for Maxwell molecules the on
collision distribution $\phi_{coll}$ is still a Gaussian):
\begin{equation}
\label{eq:fplmaxwell}
P_{FPL}^M(x)=\int \dd \bv \phi(\bv) P_{FPL}^M(x|\bv)\,\,. 
\end{equation}
This last expression can be expressed analytically in terms of Meijer
G-functions \cite{abramowitz}, but here we shall focus only on the
large length behavior, for which simpler expressions are available.
In particular, the saddle point approximation gives:
\begin{multline}
\label{eq:fplmaxlargex}
P_{FPL}^M(x) \sim \frac{\sqrt{2} \omega
  \Gamma\left(\frac{d-1}{2}\right)}{\sqrt{T} 3^{(d-1)/2}
  \Gamma\left(\frac{d}{2}\right)}\,\, \exp\left(- 3 \left(\frac{\omega
  x}{\sqrt{T}} \right)^{2/3} \right) \\ \times
  \Mfunction{_1F_1}\left(\frac{d-1}{2}\,,\frac{1}{2}\,,\frac{3}{2}
  \left(\frac{\omega x }{\sqrt{T}} \right)^{2/3} \right)\,\,,
\end{multline}
which leads to a stretched exponential behavior at large $x$ 
($\sim \exp -x^{2/3}$) 
(cf Fig. \ref{fig:fplmaxwell}).

\section{Collisional moments}
\label{app:coll}
Here we provide the expressions of the first collisional moments,
defined by Eq. (\ref{collmom}), for $\tf$ a Gaussian, and a Gaussian
multiplied by a Sonine Polynomial. We denote by
\begin{equation}
\nu_n^{(0)}=\int \dd \bv_1 v_1^n L_{\lambda}
      \frac{\ee^{-\frac{\bv_1^2}{2 T}}}{(2 \pi T)^{d/2}}\,\,,
\end{equation}
\begin{equation}
\nu_n^{(1)}=\int \dd \bv_1 v_1^n L_{\lambda}
      \frac{\ee^{-\frac{\bv_1^2}{2 T}}}{(2 \pi T)^{d/2}}
      S_2\left(\frac{v^2}{2 T} \right)\,\,.
\end{equation}
where $L_\lambda$ is the operator appearing in (\ref{eq:eigenvalue}). Hence in the Gaussian approximation one has $\nu_n \equiv
\nu^{(0)}_n$, while in the Sonine approximation $\nu_n \equiv
\nu_n^{(0)} + a_2 \nu_n^{(1)}$.  The expressions of the first
$\nu_n^{(i)}$ are:
\begin{equation}
\nu_0^{(0)}=
-\left( \frac{\left( 
        \ee^{-\lambda} - 1 \right) \,
      {\sqrt{T + 
          \Mvariable{T_0}}}}
      {
      {\sqrt{2\,\pi }}}
    \right)
\end{equation}
\begin{equation}
\nu_2^{(0)}=\frac{T^2 + 2\,T\,
     \Mvariable{T_0} + 
    2\,{\Mvariable{T_0}}^2 - 
    \ee^{\lambda}\,T\,
     \left( 3\,T + 
       2\,\Mvariable{T_0}
       \right) }{\ee^{\lambda}\,
    {\sqrt{2\,\pi }}\,
    {\sqrt{T + 
        \Mvariable{T_0}}}}
\end{equation}
\begin{widetext}
\begin{multline}
\nu_4^{(0)}=
\bigg\{{\sqrt{\pi }}\,
     \left( -\left( \ee^
           {\lambda}\,T^2\,
          \left( 8\,T^2 + 
           12\,T\,
           \Mvariable{T_0} + 
           3\,
           {\Mvariable{T_0}}^2
           \right)  \right)  +
        {\Mvariable{T_0}}^2\,
        \left( 3\,T^2 + 
          12\,T\,
           \Mvariable{T_0} + 
          8\,
           {\Mvariable{T_0}}^2
          \right)  \right)  - \\
    4\,T\,\left( T + 
       \Mvariable{T_0} \right) 
      \,\left( \frac{3\,
          \left( -1 + 
           \ee^{\lambda} \right)
           \,{\sqrt{\pi }}\,
          T\,
          \left( T + 
           \Mvariable{T_0}
           \right) }{4} + 
       \frac{{\sqrt{\pi }}\,
          \left( \ee^{\lambda}\,
           T\,
           \left( 2\,T + 
           \Mvariable{T_0}
           \right)  - 
           \Mvariable{T_0}\,
           \left( T + 
           2\,\Mvariable{T_0}
           \right)  \right) }
          {2} \right) \bigg\} \bigg/\bigg\{{
       \sqrt{2}}\,
    \ee^{\lambda}\,\pi \,
    {\left( T + 
        \Mvariable{T_0} \right)
        }^{\frac{3}{2}}\bigg\}
\end{multline}
\begin{equation}
\nu_0^{(1)}=\frac{\left( -1 + 
      \ee^{\lambda} \right) \,
    T^2}{8\,\ee^{\lambda}\,
    {\sqrt{2\,\pi }}\,
    {\left( T + 
        \Mvariable{T_0} \right)
        }^{\frac{3}{2}}}
\end{equation}
\begin{equation}
\nu_2^{(1)}=\frac{T^2\,\left( -3\,
       \left( 1 + 
         \ee^{\lambda} + 
         2\,\left( -1 + 
           \ee^{\lambda} \right)
            \right) \,T^2 - 
      \left( -6 + 
         22\,\ee^{\lambda}
         \right) \,T\,
       \Mvariable{T_0} - 
      2\,\left( -3 + 
         8\,\ee^{\lambda}
         \right) \,
       {\Mvariable{T_0}}^2
      \right) }{8\,
    \ee^{\lambda}\,
    {\sqrt{2\,\pi }}\,
    {\left( T + 
        \Mvariable{T_0} \right)
        }^{\frac{5}{2}}}
\end{equation}
\begin{multline}
\nu_4^{(1)}=\bigg\{T^2\,\left( -45\,
       \left( -1 + 
         5\,\ee^{\lambda}
         \right) \,T^4 + 
      6\,\left( 28 - 
         132\,\ee^{\lambda}
         \right) \,T^3\,
       \Mvariable{T_0} - 
      \left( -228 + 
         1000\,\ee^{\lambda}
         \right) \,T^2\,
       {\Mvariable{T_0}}^2 + 
      4\,\left( 36 - 
         128\,\ee^{\lambda}
         \right) \,T\,
       {\Mvariable{T_0}}^3 +\right.\\ \left. 
    +  8\,\left( 3 - 
         8\,\ee^{\lambda}
         \right) \,
       {\Mvariable{T_0}}^4
      \right) \bigg\}\bigg/\bigg\{8\,
    \ee^{\lambda}\,
    {\sqrt{2\,\pi }}\,
    {\left( T + 
        \Mvariable{T_0} \right)
        }^{\frac{7}{2}}\bigg\}
\end{multline}
\end{widetext}

\end{document}